\documentstyle[psfig]{mn}
\newif\ifAMStwofonts
\ifoldfss
  \ifCUPmtlplainloaded \else
    \NewTextAlphabet{textbfit} {cmbxti10} {}
    \NewTextAlphabet{textbfss} {cmssbx10} {}
    \NewMathAlphabet{mathbfit} {cmbxti10} {} 
    \NewMathAlphabet{mathbfss} {cmssbx10} {} 
  \fi
  \ifAMStwofonts
    \ifCUPmtlplainloaded \else
      \NewSymbolFont{upmath} {eurm10}
      \NewSymbolFont{AMSa} {msam10}
      \NewMathSymbol{\upi}     {0}{upmath}{19}
      \NewMathSymbol{\umu}     {0}{upmath}{16}
      \NewMathSymbol{\upartial}{0}{upmath}{40}
      \NewMathSymbol{\leqslant}{3}{AMSa}{36}
      \NewMathSymbol{\geqslant}{3}{AMSa}{3E}

    \fi
  \fi
\fi 

\ifnfssone
  \newmathalphabet{\mathit}
  \addtoversion{normal}{\mathit}{cmr}{m}{it}
  \addtoversion{bold}{\mathit}{cmr}{bx}{it}
  \newmathalphabet{\mathbfit} 
  \addtoversion{normal}{\mathbfit}{cmr}{bx}{it}
  \addtoversion{bold}{\mathbfit}{cmr}{bx}{it}
  \newmathalphabet{\mathbfss} 
  \addtoversion{normal}{\mathbfss}{cmss}{bx}{n}
  \addtoversion{bold}{\mathbfss}{cmss}{bx}{n}
  \ifAMStwofonts
    \ifCUPmtlplainloaded \else
      \UseAMStwoboldmath
      \makeatletter
      \new@mathgroup\upmath@group
      \define@mathgroup\mv@normal\upmath@group{eur}{m}{n}
      \define@mathgroup\mv@bold\upmath@group{eur}{b}{n}
      \edef\UPM{\hexnumber\upmath@group}
      \new@mathgroup\amsa@group
      \define@mathgroup\mv@normal\amsa@group{msa}{m}{n}
      \define@mathgroup\mv@bold\amsa@group{msa}{m}{n}
      \edef\AMSa{\hexnumber\amsa@group}
      \makeatother
      \mathchardef\upi="0\UPM19
      \mathchardef\umu="0\UPM16
      \mathchardef\upartial="0\UPM40
      \mathchardef\leqslant="3\AMSa36
      \mathchardef\geqslant="3\AMSa3E
    \fi
  \fi
\fi 

\ifnfsstwo
  \DeclareMathAlphabet{\mathbfit}{OT1}{cmr}{bx}{it}
  \SetMathAlphabet\mathbfit{bold}{OT1}{cmr}{bx}{it}
  \DeclareMathAlphabet{\mathbfss}{OT1}{cmss}{bx}{n}
  \SetMathAlphabet\mathbfss{bold}{OT1}{cmss}{bx}{n}
  \ifAMStwofonts
    \ifCUPmtlplainloaded \else
      \DeclareSymbolFont{UPM}{U}{eur}{m}{n}
      \SetSymbolFont{UPM}{bold}{U}{eur}{b}{n}
      \DeclareSymbolFont{AMSa}{U}{msa}{m}{n}
      \DeclareMathSymbol{\upi}{0}{UPM}{"19}
      \DeclareMathSymbol{\umu}{0}{UPM}{"16}
      \DeclareMathSymbol{\upartial}{0}{UPM}{"40}
      \DeclareMathSymbol{\leqslant}{3}{AMSa}{"36}
      \DeclareMathSymbol{\geqslant}{3}{AMSa}{"3E}
    \fi
  \fi
\fi 

\ifCUPmtlplainloaded \else
  \ifAMStwofonts \else 
    \def\upi{\pi}
    \def\umu{\mu}
    \def\upartial{\partial}
  \fi
\fi
\def\reference{\par \noindent \hangindent=0.75cm}
\def\fnsz{\footnotesize}
\def\miriad{${\sc miriad}$}
\def\deg{$^{\circ}$}
\def\kms{ km s$^{-1}$}
\def\hr{$^{\rm h}$}
\def\min{$^{\rm m}$}
\def\sec{$^{\rm s}$}
\def\amin{$'$}
\def\asec{$''$}
\def\sm{$\sim$}
\def\ha{H$\alpha$}
\def\hi{H$\sc i$}
\def\hoii{Ho$\sc ii$}
\def\sun{$_\odot$}
\newcommand{\gesim}{\,\raisebox{-0.4ex}{$\stackrel{>}{\scriptstyle\sim}$}\,}
\newcommand{\lesim}{\,\raisebox{-0.4ex}{$\stackrel{<}{\scriptstyle\sim}$}\,}
\title{High Resolution \hi\ observations of the Western Magellanic Bridge}
\author[E. Muller, L. Staveley-Smith., W. Zealey, S. Stanimirovi\'c]
  {E. Muller$^{1,2}$,L. Staveley-Smith$^2$,W. Zealey$^{1}$, S. Stanimirovi\'c$^{3}$\\
  $^1$University of Wollongong, Northfields Ave. 
  Wollongong, NSW 2500, Australia\\
  $^2$Australia Telescope National Facility, CSIRO, PO Box 76, Epping, N.S.W. 1710, Australia\\
  $^3$Arecibo Observatory, HC3 Box 53995, 
  Arecibo, Puerto Rico 00612\\}

\date{}
\pagerange{\pageref{firstpage}--\pageref{lastpage}}
\pubyear{2002}

\begin{document}

\maketitle

\label{firstpage}
\begin{abstract}
  The 21cm line emission from a 7$\times$6 degree region, east of and
  adjoining the Small Magellanic Cloud (SMC) has been observed with
  the Australia Telescope Compact Array and the Parkes telescopes.
  This region represents the westernmost part of the Magellanic
  Bridge, a gas-rich tail extending \sm14 degrees to the Large
  Magellanic Cloud (LMC).  A rich and complex neutral hydrogen
  (\hi) structure containing shells, bubbles and filaments is
  revealed. On the larger scale, the \hi\ of the Bridge is organised
  into two velocity components. This bimodality, which appears to
  originate in the SMC, converges to a single velocity component
  within the observed region.  A census of shell-like structures
  suggests a shell population with characteristics similar to that of
  the SMC. The mean kinematic age of the shells is \sm6 Myr, in
  agreement with the SMC shell population, but not with ages of OB
  clusters populating the Magellanic Bridge, which are approximately
  an factor of three older.  In general, the projected spatial
  correlation of Bridge \hi\ shells with OB associations is poor and as
  such, there does not appear to be a convincing relationship between
  the positions of OB associations and that of expanding spherical \hi\
  structures.  This survey has found only one \hi\ shell that has an
  identifiable association with a known \ha\ shell.  The origin of the
  expanding structures is therefore generally still uncertain, although
  current theories regarding their formation include
  gravitational and pressure instabilities, HVC collisions and ram
  pressure effects.

\end{abstract}


\section{Introduction}
The Magellanic Bridge is a loosely defined column of gas, comprising
mostly neutral hydrogen, found between the Small and Large Magellanic
Clouds (SMC and LMC respectively).  The Bridge was discovered
originally through 21cm observations by Hindman et al. (1961), and has
been mapped in the \hi\ line at increased spatial resolution by
Mathewson, Cleary \& Murray (1974), and at increased velocity
resolution by McGee \& Newton (1986).  The most recent \hi\
observations are presented by Putman (1998), and Br\"uns, Kerp
\& Staveley-Smith (2000).

The tidal influence of the Magellanic Clouds on each other is widely
considered to be the mechanism responsible for the development of the
Magellanic Bridge (eg. Putman 2000; Demers \& Battinelli 1998;
Staveley-Smith et al. 1998), and has been modelled as such through
numerical simulations (eg. Gardiner, Sawa \& Fujimoto 1994; Gardiner
\& Noguchi 1996; and Sawa, Fujimoto \& Kumai 1999).  The simulations
suggest that formation of the Bridge may have begun during the most
recent of a series of close Cloud/Cloud interactions, around 200 Myr
ago.  Zaritsky et al. (2000) suggest that the SMC may also have been
subject to a period of ram pressure, and have measured a shift in the
centre of the young blue population relative to that of the older
population. The degree to which this hydrodynamic effect has
influenced the evolution of the Magellanic System has not yet been
quantified.
 
Studies of the morphology of the \hi\ in the SMC have been made by a
few groups: \hi\ shells have been identified and catalogued by
Staveley-Smith et al. (1997) and by Stanimirovi\'c et al. (1999). The
statistical properties of the Interstellar Medium (ISM) have been
studied by Stanimirovi\'c et al. (1999), Stanimirovi\'c (2000) and
Stanimirovi\'c \& Lazarian (2001).  The shell population, its evolution
and relationship with star forming regions is studied by Oey \& Clarke
(1997).  These statistical studies lead to comparisons that can assist
in the understanding of the shell evolutionary environment for these
systems.  Studies of the \hi\ shell population in other galaxies have
been used as a probe into the physical processes active in the local
ISM; Puche et al.  (1992) compiled a catalogue of the shells in the
Magellanic type galaxy Holmberg II (\hoii), while Walter \& Brinks
(2001) have made similar observations of the shell population of the
Magellanic type DDO 47 galaxy.  Wilcots \& Miller (1998) made a
high-resolution study of the \hi\ in the dwarf irregular galaxy IC10,
and found that it is dominated by a rather chaotic and frothy \hi\
morphology.

Further observations and analysis of the \hi\ shell population of \hoii\
designed to test the standard 'stellar wind' mechanism of shell
generation (Rhode et al. 1999) have shown that there are observational
inconsistencies with this idea. Alternative shell mechanisms, such as
gamma ray bursts and high velocity cloud (HVC) impacts have been
proposed by these authors. Other suggestions for the formation of
these \hi\ holes by ram pressure have been made by Bureau \& Carnigan
(2002).
An examination of the shell population of the Galaxy (Ehlerova \&
Palous, 1996) suggests that the formation of these shells is more
likely to be the product of stellar wind, rather than collisions with
HVCs.
 
We present here high spatial and velocity resolution observations of
the Magellanic Bridge, conducted with both the Australia Telescope
Compact Array (ATCA) and with the Parkes telescope \footnote{The
  Australia Telescope Compact Array and Parkes telescopes are part of
  the Australia Telescope which is funded by the Commonwealth of
  Australia for operation as a National Facility managed by CSIRO}.
The combined data cube is sensitive to structure on all angular scales
between 98 arcsec and \sm6 degrees.  A survey for shells and bubbles
in the observed volume has been made, and statistical analysis is
compared with that of the SMC and with the \hoii\ shell survey by
Puche et al. (1992).

\vspace{5mm}

Section~\ref{sec:atcaobs} and Section~\ref{sec:parkesobs} outline the
procedures involved in the observations made with the ATCA and with
the Parkes telescope, while Section~\ref{sec:datamerge} outlines the
methods used in merging these two datasets.
Section~\ref{sec:cuberesults} discusses the general appearance and
highlights some of the more dramatic features of the \hi\ cube. The
shell selection criteria are defined in Section~\ref{sec:shellsurv}.
Section~\ref{sec:shellanalysis} contains a statistical summary of the
results and in Section~\ref{sec:corr} we present a study of the
correlation of the shell and OB association population of the Bridge.
In Section~\ref{sec:ha} we compare an \hi\ shell found from this study
with an \ha\ region found already within the Bridge.  Limitations
affecting the shell survey are discussed in Section~\ref{sec:limits}.
Section~\ref{sec:discussion} contains a discussion of the stellar wind
model, as applied to the Magellanic Bridge shell population, as well
as a comparison with the energetics of the SMC shell
population. Alternative shell generation mechanisms are discussed in
Section~\ref{sec:alternatives}, and we summarise our findings in
Section~\ref{sec:summary}.

\section{Observations}

\subsection{ATCA Observations}\label{sec:atcaobs}

\begin{figure}
  \centerline{\psfig{file=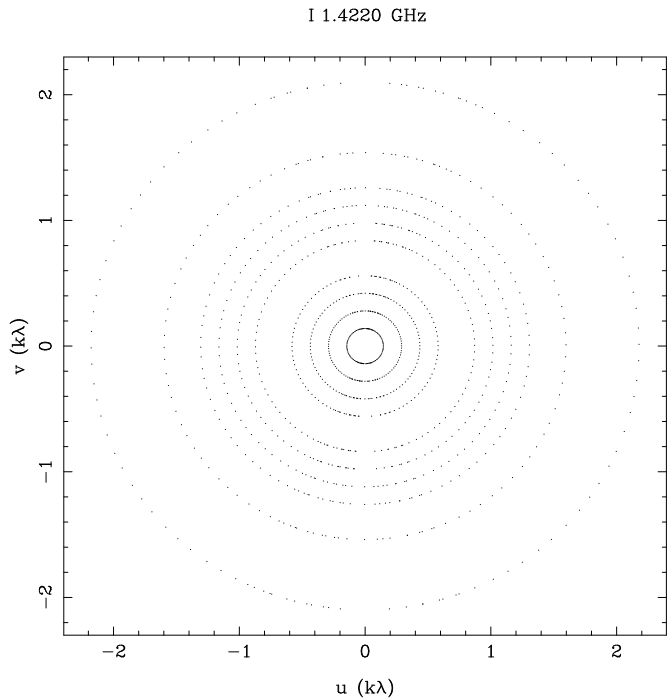}}
\caption{UV coverage for pointing \#20 of 'block 4', centred at
  RA 02\hr06\min07.6\sec, Dec$-$74\deg39\amin55.1\asec. This is a typical
  example of the UV coverage for these observations.  There are a
  total of 480 centres over the entire field.}
\label{fig:UV}
\end{figure}

Observations of the 21cm \hi\ line were made over a 7x6 degree field
using the 375 m configuration of the ATCA.  These observations were
made over three sessions: 1997 April 13, 15-16, 18; 1997 October 9-15;
and 2000 January 29-February 2. The 7x6 degree field is broken into 12
'blocks', with each block containing 40 pointings. Each pointing is
visited for twenty seconds, approximately once every 15 minutes.  The
total integrated time for each pointing is thus \sm 16 minutes.

Most of the blocks were observed at this rate over 12 hours for
complete UV sampling. Five of the blocks were not observed for a full
12 hours, although these are incomplete by only 6 per cent, and it is not
considered that significant artifacts exist in the final image data.
The UV coverage of the central pointing (pointing \#20) for block 4 is
shown in Fig.\ref{fig:UV} as an example of the UV sampling for these
observations.

The ATCA observations were made using a 4 MHz bandwidth, with 1024
channels at a central frequency of 1.420 GHz, resulting in a channel
spacing of 0.83\kms\ before Hanning smoothing.

The calibrator PKS B1934-638 was observed as a flux standard.  This is
assumed to have 14.9 Jy at 1.42 GHz. Where this could not be observed,
PKS B0407-658 was observed instead, this is
assumed to have 14.4 Jy at 1.42 GHz.  PKS B0252-712 was used for phase
calibration where possible, otherwise PKS B0454-810 was used. At 1.42
GHz, PKS B0252-712 is scaled to 5.7 Jy, and PKS B0454-810 is scaled to
1.10 Jy.

The primary flux calibrator was observed before (and after, where
possible) each observing session, none of which lasted for more than
twelve hours. The phase calibrators were observed approximately every
fifteen minutes.

The \miriad\ data reduction suite was exclusively used for processing
and reduction of the ATCA visibility data, and for the construction of
the final image data cube. The data were Hanning smoothed using the
\miriad\ task {\sc atlod} to a channel spacing of \sm1.63 \kms. Standard
editing and calibration methods were followed to create an image
datacube (eg. Stanimirovi\'c, 1999): {\sc invert} was used to linearly create
a dirty image mosaic from the visibilities, using a robustness
parameter of zero to down-weight the longer baselines, this has a
 final sensitivity of 0.9 K. Deconvolution was done with
{\sc mosmem}, which uses a maximum entropy algorithm to deconvolve the dirty
image cube.  {\sc restor} was used to add back in the residuals and convolve
the data with a 98\asec gaussian function.

\subsection{Parkes Observations}\label{sec:parkesobs}
Observations were made using the Multibeam receiver on the 64m Parkes
telescope, during 1999 November 2nd-8th. Only the seven inner
receivers of the Multibeam array were used, with each beam having a
FWHM width of 14.1\amin. Forty-eight overlapping scans were made in
Declination, using the on-the-fly mapping mode at 1\deg/min in
Declination. The scans were centred on RA 02\hr00\min, Dec$-$72\deg
20\amin, and extended 8\deg in Declination and in RA, large enough to
fully encompass the area observed with the ATCA.  The scans were
interleaved with a spacing of 1.11\deg, with a continuously rotating
receiver so as to maintain a relative angle of the scan tracks of
19.1\deg to the sky. The final spacing between adjacent beam tracks
was \sm6.7\amin. An 8 MHz bandwidth was used, with 2048 channels and
centred on 1.42 GHz.  This gave a channel spacing of 0.83\kms.  These
observations were frequency-switched with a frequency throw of
+3.5MHz, equivalent to \sm 896 channels and \sm739.5 \kms.

The data were reduced and bandpass calibrated using the ${\sc aips++}$
online data reduction system {\sc livedata}.  {\sc livedat} was also
used to apply velocity corrections. The cube was gridded using the
{\sc sdfits2cube} algorithm ({\sc slap} package, Staveley-Smith, L.
Priv comm.), resulting in a beam FWHM of 15.7\asec.  The final Parkes
data cube encompassed Heliocentric velocities from 100\kms\ to 350\kms.

\subsection{Merging of ATCA and Parkes Data}\label{sec:datamerge}
Stanimirovi\'c (1999) concluded that for telescopes where there is
significant overlap in the UV plane (i.e. D$_{single dish}>>$ shortest
baseline), the differences between merging methods, where single dish
data is added to interferometric data before, during, or after
deconvolution are usually minimal, and as such, a linear,
post-deconvolution method was employed here for its simplicity. The
ATCA and Parkes data cubes were combined using the \miriad\ task {\sc immerge}. The Parkes data were first Hanning smoothed to the same
velocity resolution as the ATCA data and regridded to the same spatial
and velocity dimensions using the \miriad\ task {\sc regrid}.  {\sc immerge} scales the Parkes data by comparing the real and imaginary
parts of the ATCA and Parkes data in a region in the Fourier plane
that is common to both datasets.  It then linearly adds the two data sets
so the combined amplitude-spatial frequency
response curve returns to a gaussian form, with a width equal to that
of the ATCA data (\miriad\ manual, Sault, Killeen 1999). For perfectly
calibrated and stable telescopes the scaling factor should be
equal to 1, however, data quality varies over time and from telescope
to telescope, and this factor is determined more accurately using
${\sc immerge}$.  A scaling factor of \sm1.15 for the Multibeam
dataset is used during combination of the Parkes data from these
observations.

The combined cube is converted to brightness temperature using the
relation; $S=2k{\Omega}T_B/\lambda^2$, where $\Omega$ is the beam
area of the combined cube (=$\Omega_{ATCA}$). The combined cube covers
a velocity range of 100\kms\ to 350\kms\ with 152 channels.  It
has a velocity channel spacing of 1.65\kms\ and an RMS of 0.8 K
beam$^{-1}$ as measured in line free channels of the cube.  This
corresponds to a column density of 1.7$\times$10$^{18}$cm$^2$ for each
velocity channel, assuming the mass is optically thin. The final
angular resolution is 98\asec.

Fig.\ref{fig:ppix1} shows the integrated intensity maps of the RA-Dec
and RA-Velocity projections of the combined ATCA/Parkes data cube,
while the Velocity-Declination projection are shown in Fig.
\ref{fig:ppix2}

The RA-Dec projection of the cube, shown in Fig.\ref{fig:ppix1}a,
reveals the Magellanic Bridge is dominated by filaments, loops and
arcs, down to the smallest scale of 98\asec.  This map shows a region
of relatively high \hi\ column density running East-West, where the \hi\
column density is higher than the background diffuse component by a
factor of \sm20. The highest column density in this east-west strip is
5.5$\times$10$^{21}$cm$^{-2}$ at RA 1\hr23\min59\sec, Dec $-$73\deg07\amin43\asec, although this point is inside the SMC.  Outside
of the SMC eastern wing, the highest column density is
\sm2.8$\times$10$^{21}$cm$^{-2}$ at RA 01\hr58\min09\sec,
Dec $-$74\deg17\amin28\asec.

\subsection{The complete \hi\ cube.}\label{sec:cuberesults}
\begin{figure*}
\caption{Integrated intensity maps of Right Ascension-Declination and Right
  Ascension-Velocity projections of combined ATCA-Parkes datacube.
  Greyscale is linear as shown on the intensity wedge. Units are
  K$^.$\kms. A conversion to column density (atm/cm$^{-2}$) can be
  made by multiplying the integrated intensity by
  1.8$\times$10$^{18}$.  Shown here are a)RA-Dec projection and
  b)Vel-Dec projection (see Fig.\ref{fig:ppix2} for Ra-Vel
  projection) Velocities are in the Heliocentric rest frame. The upper
  and lower lines in b) denote velocities of 38 and 8 \kms\
  respectively, relative to Galactic centre.} \centerline{
  \psfig{file=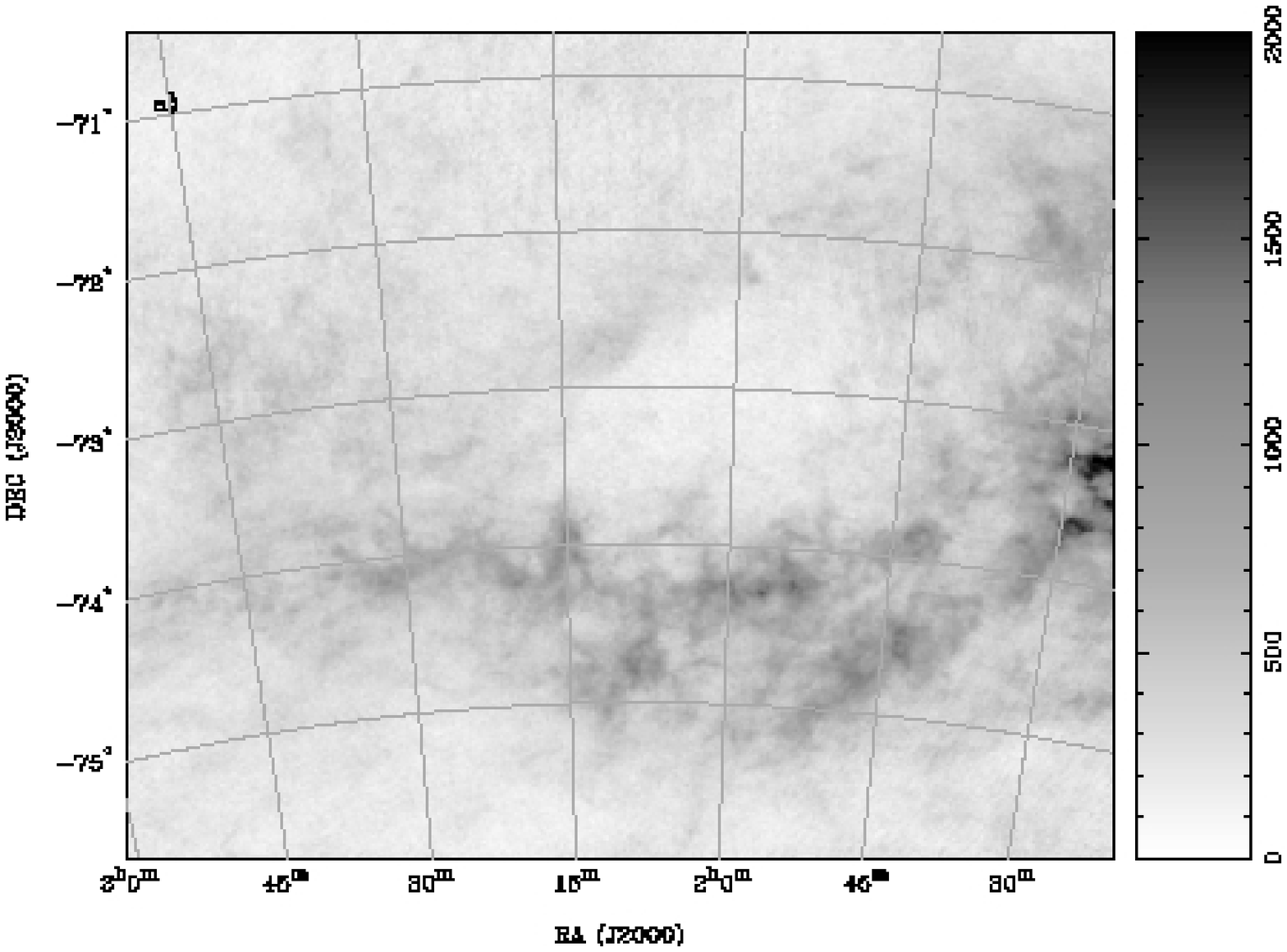,width=17.5cm}} 
  \centerline{
  \hspace{6mm} 
  \psfig{file=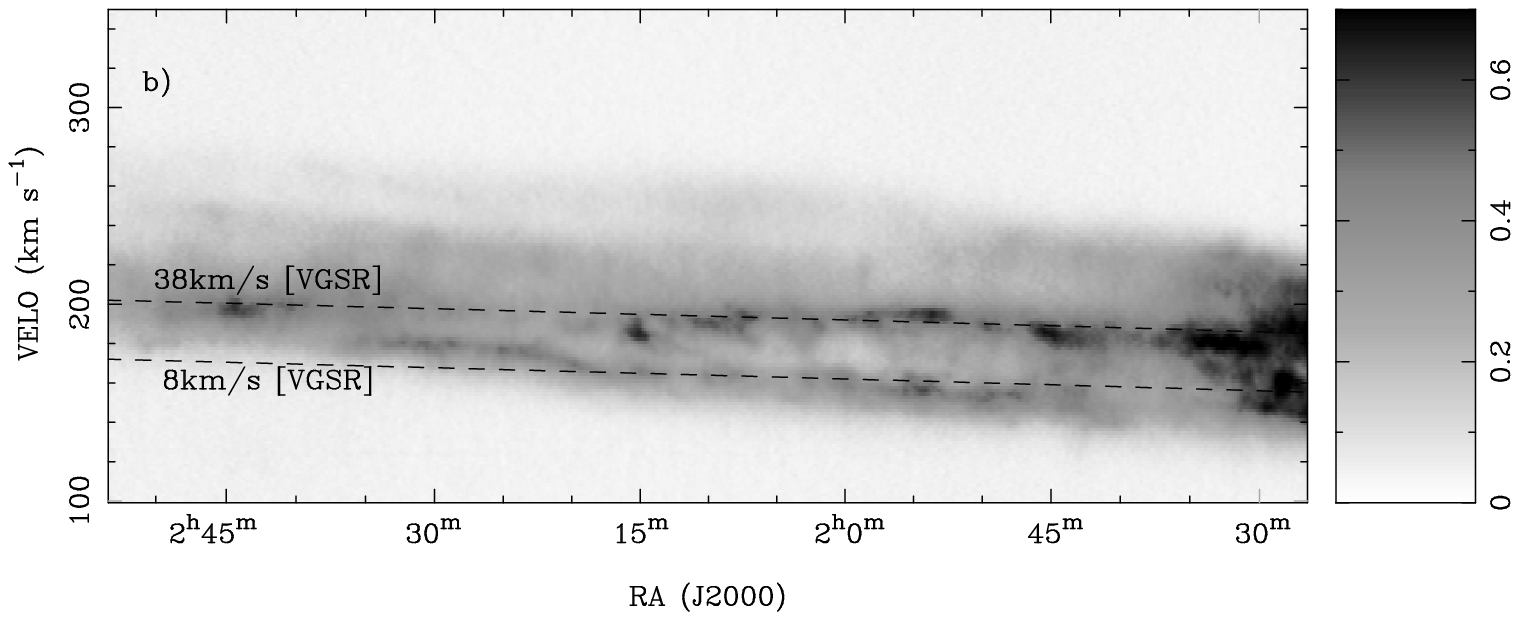}}
\label{fig:ppix1}
\end{figure*}

\begin{figure}
\caption{Integrated intensity maps of Velocity-Declination projection of
  combined ATCA-Parkes datacube, units and scaling are as for
  Fig.\ref{fig:ppix1}}
 \centerline{\psfig{file=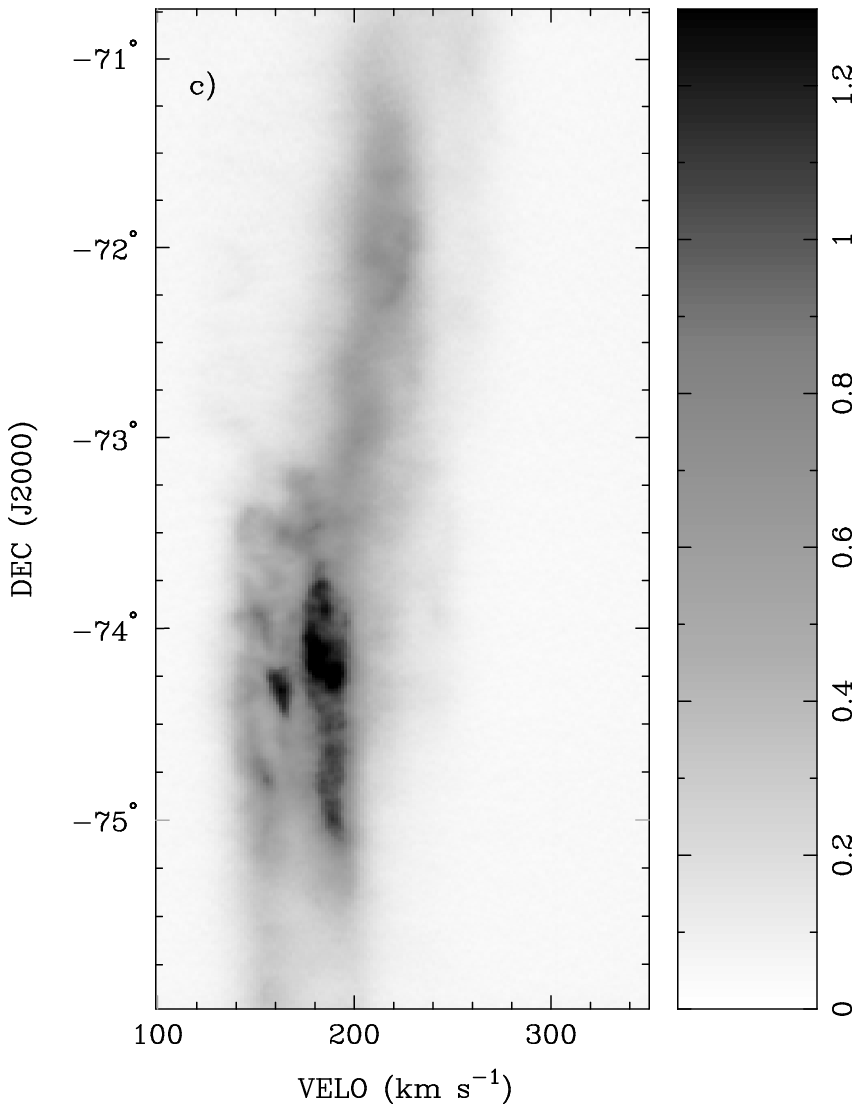}}
\label{fig:ppix2}
\end{figure}

Perhaps the most striking feature of these maps is a large loop shaped
filament, centred on RA \sm02\hr 09\min 59\sec, Dec $-$73\deg
21\amin56\asec. The column density of the loop varies
between \sm0.5-1$\times$10$^{21}$cm$^{-2}$, while the mean column
density of the interior of the loop is
\sm0.2$\times$10$^{21}$cm$^{-2}$. The radius of the loop is \sm1\deg,
corresponding to a projected diameter of \sm1.1 kpc, assuming that its
distance is the same as the adjacent SMC of 60 kpc (e.g. Stanimirovi\'c
1999).

Lines of constant velocity (38\kms\ and 8\kms) in the galactic rest
frame are included on the RA-Velocity map of Fig.\ref{fig:ppix1}b.
The conversion from Heliocentric to Galactic rest-frames is made with
the relation:
\noindent
$V_{GSR}=V_{Hel}+232\sin(l)\cos(b)+9\cos(l)\cos(b)+7\sin(b)$\\ 
\noindent(eg. Paturel et al, 1997).

The velocity projections of the cube are shown in
Figs.\ref{fig:ppix1}b and \ref{fig:ppix2}. These figures reveal more
striking structure, in particular, an apparent velocity bimodality can
be seen in the RA-Vel plot of Fig.\ref{fig:ppix1}b, and to a lesser
extent in Fig.\ref{fig:ppix2}.  Closer to the SMC, the velocity
profile appears to be trimodal, and this will be discussed later. The
separation of the velocity peaks is \sm30-40\kms\ and appears to be
roughly constant in RA from the SMC up to RA 2\hr17\min, where it
suddenly converges to a single velocity of 180\kms\ [Helio].
Fig.\ref{fig:ppix2} also reveals interesting velocity structure. The
bimodality seen in Fig.\ref{fig:ppix1}b manifests as two parallel
sheets in velocity at \sm150\kms and \sm190\kms. In addition, a large mass, contained between
\lesim$-$71\deg to \sm $-$73\deg30\amin, can be seen to have a
significantly higher positive velocity than the bulk of the Bridge gas
by \sm+40\kms.  The large ring-shaped \hi\ void mentioned above is
found within this higher velocity mass.  Intensity maps of the cube,
integrated over groups of five channels (intervals of \sm8 \kms) are
shown in Fig.\ref{fig:chanmap}.  The SMC appears in the first few
frames on the western side and in general, it can be seen that in the
Heliocentric rest frame, the positive velocity of the \hi\ of the
Magellanic Bridge increases with Right Ascension. The large loop
filament is particularly obvious in the higher velocity frames centred
on V$_{Hel}$=194-227 \kms.

The mass of the observed region excluding the SMC in the western edge
of the observed area (west of \sm1\hr35\min), and a 35\asec margin
around the edge of the image is \sm1.5$\times$10$^{8}$ M\sun.
However, Fig~\ref{fig:ppix1}b shows that the central region contains
the most mass, and we will see later that the the expanding shell
population appears to be more prevalent in this area. If we examine
only the central higher-column density region, bounded by $-$75.5\deg to
$-$73.5\deg and 1\hr34\amin to 2\hr46\amin (corresponding to a height of
\sm2.1 kpc, and a length of \sm5 kpc, see Fig.\ref{fig:ppix1}a), we find
an enclosed mass of approximately 7.4$\times$10$^{7}$ M\sun\ with a
surface mass distribution of \sm7 M\sun\ pc$^{-2}$. There are two
reasonable approaches to calculate the density of this region: 
\begin{enumerate}
\item[1.] We can assume that the mass of this region is contained
  within a cylinder of radius 2.1 kpc/2 = 1.1 kpc and height 5 kpc, then
  the approximate volume density is n$_o$\sm0.2 cm$^{-3}$.

\item[2.] Demers \& Battinelli (1998) inferred a depth of the Bridge
  of \sm 5 kpc by measuring a difference in distance modulus of two
  adjacent (separated by \sm7') O associations.  If the high density
  region of the Bridge is modelled as a slab, with a width of 5 kpc, a
  length of \sm5 kpc and a thickness of \sm2.1 kpc, the number density
  of \hi\ is \sm0.06 cm$^{-3}$.
\end{enumerate}
These figures are derived from the central high-density region and the
estimate will represent a maximum value for the observed area. The
more tenuous gas north and south of the central region has column
densities \sm0.3-0.5 of the central high density region.  As the
latter value of n=0.06 cm$^{-3}$ is derived using real depth
measurements, we use this value throughout this study.

\begin{figure*}

  \centerline{\psfig{file=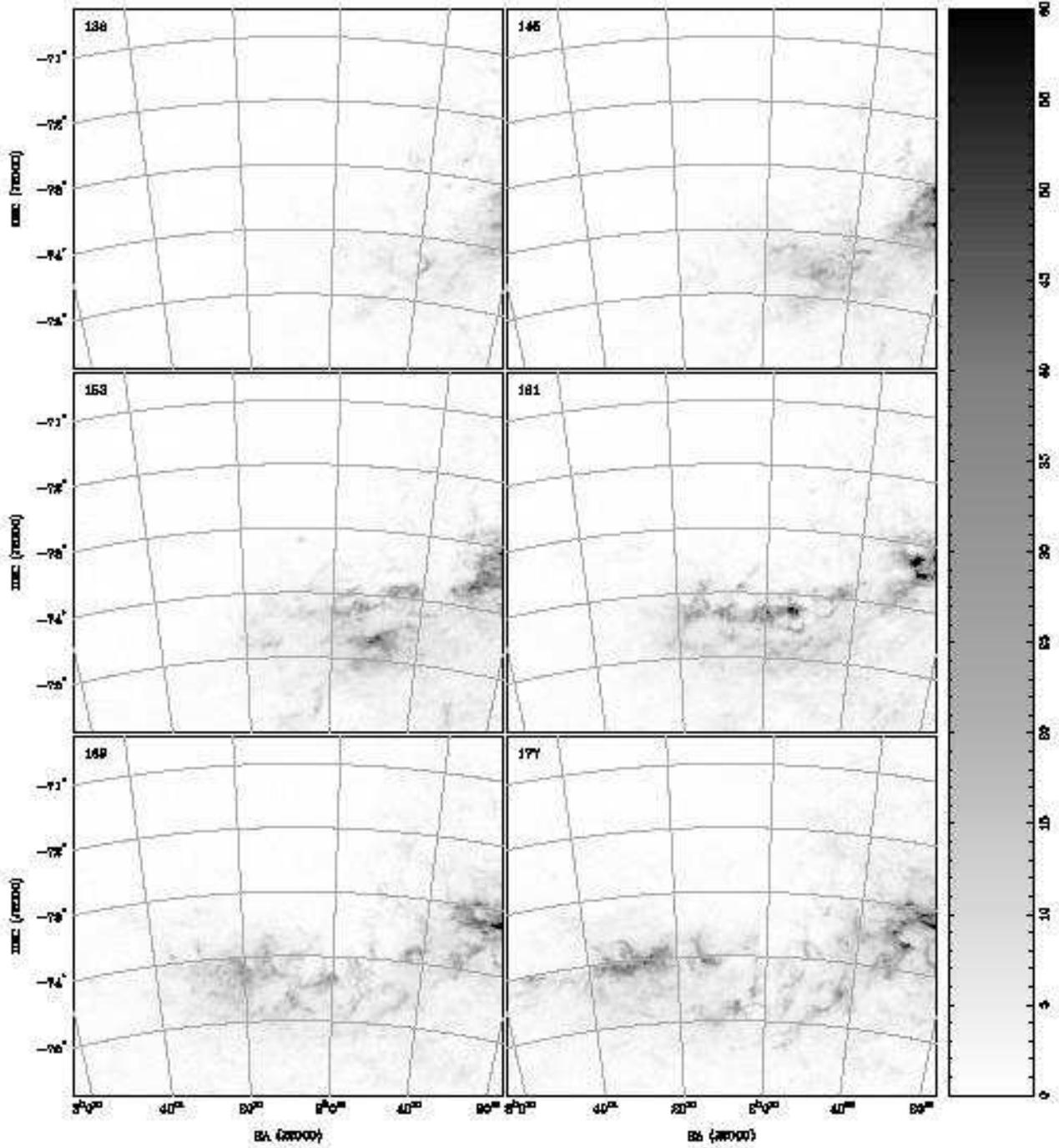,width=17.5cm}}
\caption{Composite ATCA-Parkes \hi\ channel maps over velocity
  range containing significant signal, of 135\kms\ to
  181\kms\ [Helio].  Each of these panels are integrated over groups
  of 5 velocity channels (\sm8 \kms). The central velocity is in the
  top left of each map, and the greyscale is a linear transfer
  function, with units in K}
\label{fig:chanmap}
\end{figure*}
\eject 
\addtocounter{figure}{-1}
\begin{figure*}
  \centerline{\psfig{file=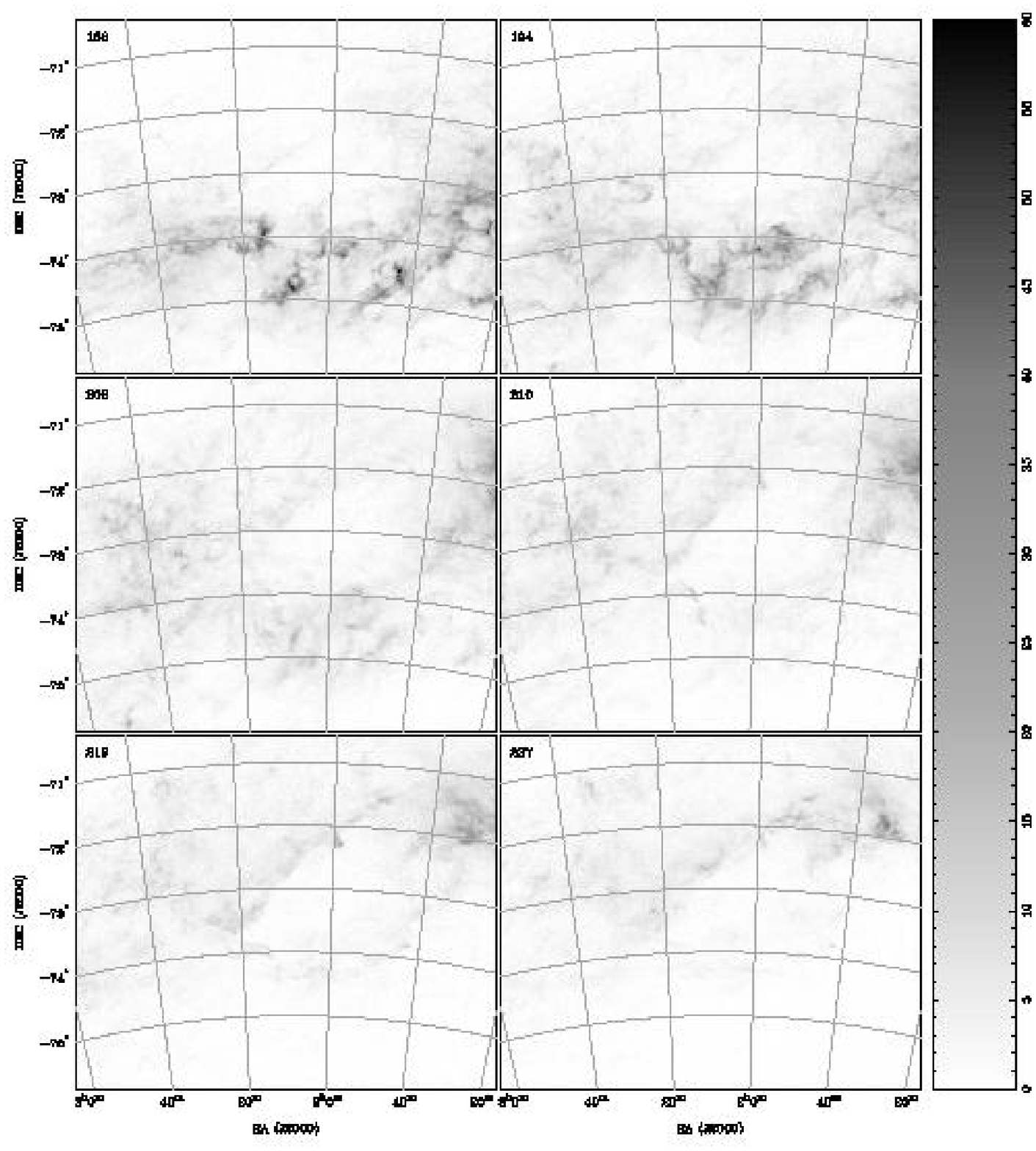,width=17.5cm}}
\caption{Continued, covering the velocity range
  \sm182-234\kms.  The large \sm1 kpc loop features prominently in this
  range.}
\end{figure*}

\section{The Shell Survey}\label{sec:shellsurv}

A shell or bubble within a gas cloud can be generated through a
variety of mechanisms. Currently, one of the most popular theories
involves a young energetic star or cluster, which ionises the
surrounding gas into a hot, high pressure region and produces a
spherical, high-density shockwave which propagates into the ambient
neutral gas (Shu 1992). In addition to ionisation, the young star
constantly sheds mass in a stellar wind which impinges on the local
gas. The result is a relatively low density sphere, enclosed by a
higher density hydrogen shell, the outer edge of which is neutral and
moves at supersonic velocities into the ambient gas (McCray \& Snow
1979).  Supernova events will also deposit energy into the medium
(Cox 1972) and alternative mechanisms such as Gamma ray bursts
(Efremov , Elmegreen and Hodge, 1998) and HVC-disk collisions
(Tenorio-Tagle 1981 and Tenorio-Tagle et al. 1987) may also produce
elliptical structures with similar appearance.  In general, it is very
difficult to determine the process by which a particular shell has been
formed. \hi\ column density maps will show embedded expanding shells as
a relatively bright ring in spatial co-ordinates, and depending on the
expansion velocity of the shell, as an ellipse in position-velocity
space.

All projections (RA-Dec, RA-Vel and Vel-Dec) of the ATCA/Parkes
composite data cube were examined for expanding shells, using the
{\sc karma} applications {\sc kslice-3d} and {\sc kpvslice}.  The criteria for this
survey are based on those defined by Brinks \& Bajaja (1986) and Puche
et al. (1992).

A ring feature was catalogued as an expanding \hi\ shell if the
following criteria were satisfied:

\begin{description}
\item[\it{i} -]An expanding shell must be observed as a complete ring,
  or rough ring shape, within the velocity range occupied by the shell
  (Criterion \textit{iv}, Puche et al., 1992).

\item[\it{ii} -]Expansion must be observable in both Position-velocity
  projections across at least three velocity channels, and with a
  stationary centre throughout the channel range occupied by the
  shell.  This criterion was modified from Criterion \textit{ii},
  Puche et al, where the ring integrity is examined only in the RA-Dec
  projection

\item[\it{iii} -]The rim of the ring has good contrast (i.e.
  relatively high column density) with respect to the ambient column
  density of the channel maps (Criterion \textit{iii}, Puche et al., 1992).
\end{description}

Note that the criteria here target rim-brightened expanding \hi\
shells, and attempt to exclude \hi\ holes that do not appear to show
expansion in both velocity projections.  This differs from the \hi\
holes studied in IC10 by Wilcots \& Miller (1998) at a channel spacing
of \sm2.9 \kms, where although all \hi\ holes were examined for a
double peaked velocity profile, none were found.  The velocities of
the receding and approaching sides of the shell, as well as the shell
radius (in arcmin), were measured with the {\sc karma} application
{\sc kshell}.
The Heliocentric velocity of the shell was calculated as the average
of the velocities of the approaching and receding sides of the shell,
while the expansion velocity is half the absolute difference in these
velocities.  It should be pointed out here that in an effort to reduce
an element of subjectivity, typically inherent in surveys for \hi\
expanding shells, it was required that the three above criteria be
strictly satisfied.  The effects of such strictness are apparent in a
statistical examination of the resulting dataset, and these are
discussed in Section~\ref{sec:limits}.

Figs~\ref{fig:ppix1}a and \ref{fig:ppix1}b show that the Magellanic
Bridge connects smoothly with the SMC, both spatially, and in
velocity.  These figures also show that the orientation of the Bridge
is quite parallel to the lines of constant Declination.  By
assuming a distance of 60 kpc to the centre of the Small Magellanic
Cloud at RA=1.0\hr and a distance of 50 kpc to the centre of the LMC at
RA=5.33 hours, we estimate the distance to individual shells within
the Magellanic Bridge with a simple linear interpolation between the
two Clouds from an empirical relation: $D
( kpc)=57.7-(RA-2^{h})\times2.3$.

To compare shell kinematic ages and luminosities between the
Magellanic Bridge, the SMC and other \hi\ systems, the following
relations, derived by Weaver et al. (1977), are used.  The shell
kinematic age is found from $T_s =
\frac{3}{5}\biggl(\frac{R_s}{v_{exp}}\biggr)$, with R$_s$ and
V$_{exp}$ the shell radius in parsecs and shell expansion velocity in
\kms\ respectively.  The shell luminosity refers to the power deposited
into the local medium through the action of the stellar wind, and is
given by
$L_s=1.5\times10^5\biggl(\frac{r}{100pc}\biggr)^5\biggl(\frac{T}{10^6yr}\biggr)^{-3}\biggl(\frac{n_o}{1
  cm^{-3}}\biggr)L_{\odot}$.

Using a solar luminosity\sm3.9$\times$10$^{33}$ erg sec$^{-1}$, the
relation for shell luminosity is further multiplied by the dynamic age
(in seconds) arrive at an estimate of the total shell energy.  This
formula is derived for a continuous injection of energy into the
shell, does not take into account any other external effects,
including magnetic and gravitational forces, and assumes a perfectly
homogeneous local gas environment.  We use n=0.06cm$^{-3}$ as
estimated in Section~\ref{sec:cuberesults}.

The shell radii, expansion velocities, Heliocentric velocities and
dynamic ages are plotted as a function of Right Ascension, and
collated for a statistical analysis and comparison with the shells
within the SMC.

\subsection{\hi\ shell survey results and analysis}\label{sec:shellanalysis}
We have catalogued 163 candidate shells according to the selection
criteria of Section \ref{sec:shellsurv}. The RA, Dec, heliocentric
velocity, expansion velocity, radius (in parsec), kinematic age and
energy for each shell are shown in Table~\ref{tab:Shelltab}.

\begin{figure*}
  \centerline{\psfig{file=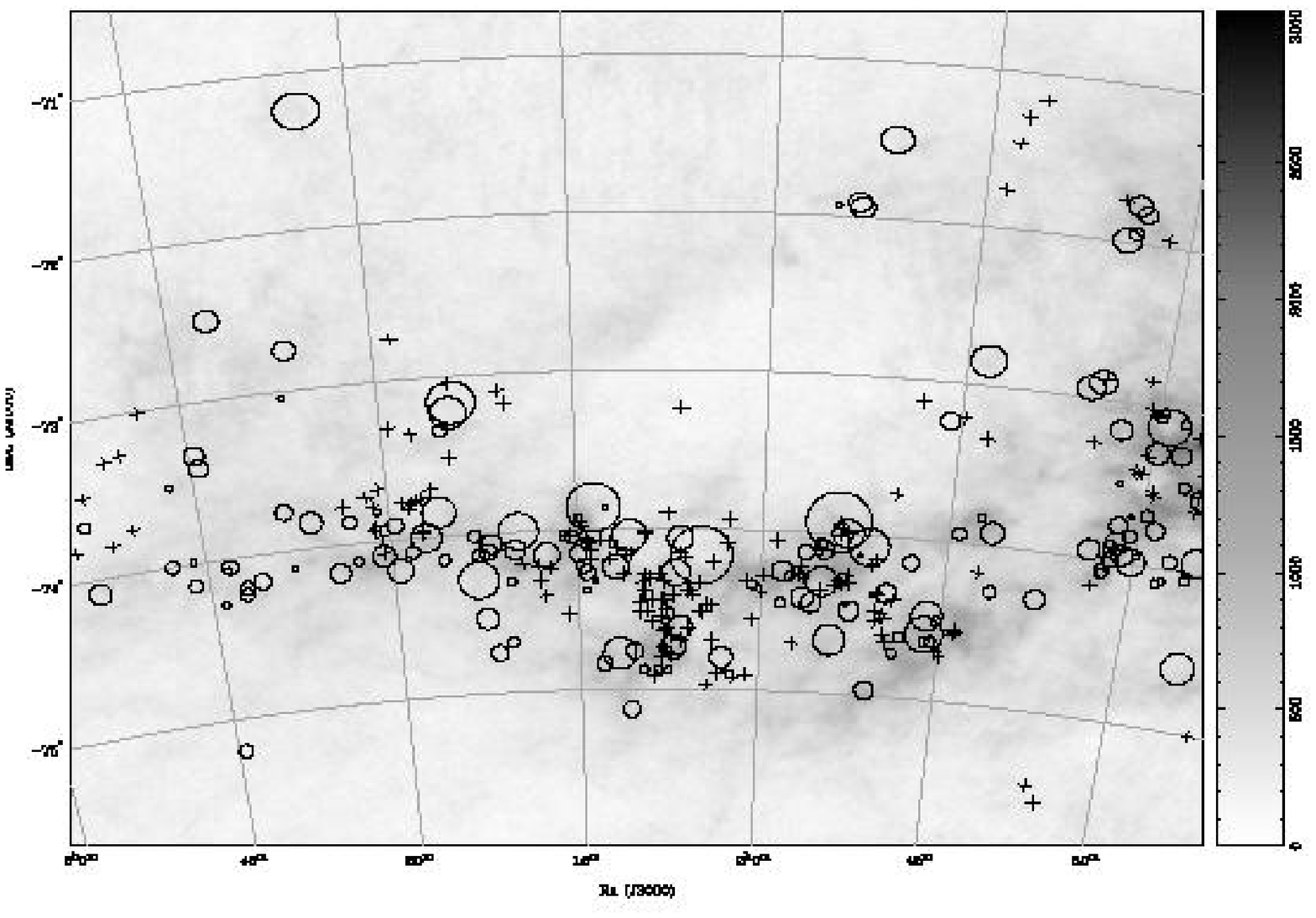,width=19cm}}
\caption{
  \hi\ shells and OB associations within the Western Magellanic Bridge
  from this survey are overlaid on an integrated intensity \hi\ image.
  The positions of shells are represented as circles, where the circle
  radii correspond to shell radii.  The shells appear to be mainly
  confined to regions of higher \hi\ column density.  Positions of
  Young OB associations (crosses) have been taken from a catalogue by
  Bica \& Schmitt (1995).  Transfer function is linear with units in
  K$^.$kms.}
\label{fig:shellolay}
\end{figure*}

The positions of the surveyed shells, as well as the positions of OB
associations from a catalogue compiled by Bica \& Schmitt (1995), are
overlaid on an integrated intensity map of the Magellanic Bridge in
Fig.\ref{fig:shellolay}. A visual inspection of this figure suggests
a good spatial correlation between \hi\ column density and expanding
shells, and OB associations.  This is discussed further in
Section~\ref{sec:corr}.

\begin{figure}
  \centerline{\psfig{file=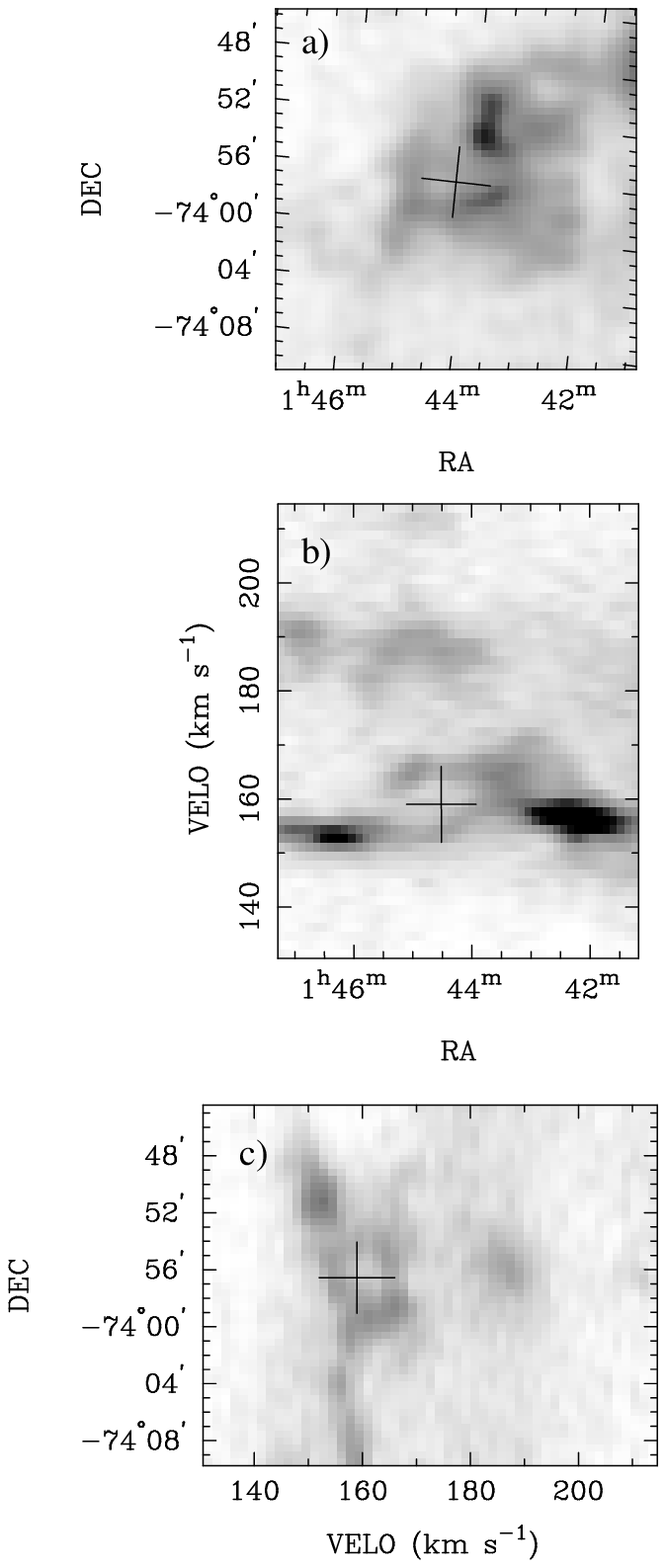}}
\caption{Three cuts through the combined datacube centred on shell
  \#51.  {\em Top} RA-Declination projection, {\em Middle} RA-Velocity
  projection, {\em Bottom} Velocity-Declination projection.  The cross
  overlay shows the position size and velocity of the shell as it
  appears in the catalogue of Table\ref{tab:Shelltab}. The greyscale
  is a linear transfer function, ranging from \sm0 to \sm 50 Kelvin.}
\label{fig:example}
\end{figure}

Fig.\ref{fig:example} shows the RA-Dec, RA-Vel and Vel-Dec projections
for an example shell \#51. The expanding volume is clear in the
figure.

\begin{table*}
\begin{center}
\caption{Table of Magellanic Bridge Shell Parameters}
\label{tab:Shelltab}
\vspace{5mm}
\begin{tabular}{lcccccccc}\hline\hline
\fnsz{Shell}&
\fnsz{Right Ascension}& 
\fnsz{Declination}& 
\fnsz{Heliocentric}& 
\fnsz{Exp. Vel.}& 
\fnsz{Radius}&
\fnsz{Radius}&
\fnsz{Dynamic}&
\fnsz{Energy}\\
\fnsz{Number}&
\fnsz{(J2000)}&
\fnsz{(J2000)}&
\fnsz{Vel. (\kms)}&
\fnsz{(\kms)}&
\fnsz{($'$)}&
\fnsz{(pc)}&
\fnsz{Age (Myr)}&
\fnsz{(log erg)}\\
\hline\hline
\fnsz{  1}&\fnsz{01:21:36.1}&\fnsz{-73:29: 3.0}&\fnsz{150.3}&\fnsz{ 4.1}&\fnsz{ 3.0}&\fnsz{ 54.3}&\fnsz{ 7.9}&\fnsz{47.9}\\
\fnsz{  2}&\fnsz{01:22:39.7}&\fnsz{-73: 5:30.2}&\fnsz{172.6}&\fnsz{12.1}&\fnsz{ 5.0}&\fnsz{ 90.7}&\fnsz{ 4.5}&\fnsz{49.5}\\
\fnsz{  3}&\fnsz{01:22:59.0}&\fnsz{-73:25:30.9}&\fnsz{168.2}&\fnsz{11.0}&\fnsz{ 3.5}&\fnsz{ 63.1}&\fnsz{ 3.4}&\fnsz{49.0}\\
\fnsz{  4}&\fnsz{01:23: 1.1}&\fnsz{-74: 6:23.3}&\fnsz{158.5}&\fnsz{ 7.6}&\fnsz{ 3.0}&\fnsz{ 54.7}&\fnsz{ 4.3}&\fnsz{48.5}\\
\fnsz{  5}&\fnsz{01:23: 6.2}&\fnsz{-73:30:24.0}&\fnsz{189.5}&\fnsz{ 2.1}&\fnsz{ 2.9}&\fnsz{ 53.3}&\fnsz{15.0}&\fnsz{47.3}\\
\fnsz{  6}&\fnsz{01:24:24.5}&\fnsz{-73:23:24.9}&\fnsz{151.8}&\fnsz{ 8.0}&\fnsz{ 3.2}&\fnsz{ 58.3}&\fnsz{ 4.4}&\fnsz{48.6}\\
\fnsz{  7}&\fnsz{01:24:48.5}&\fnsz{-74:35:50.9}&\fnsz{192.9}&\fnsz{ 3.8}&\fnsz{ 5.9}&\fnsz{107.4}&\fnsz{17.1}&\fnsz{48.7}\\
\fnsz{  8}&\fnsz{01:24:58.6}&\fnsz{-73:11:38.8}&\fnsz{147.2}&\fnsz{ 6.0}&\fnsz{ 1.7}&\fnsz{ 30.8}&\fnsz{ 3.1}&\fnsz{47.5}\\
\fnsz{  9}&\fnsz{01:24:59.9}&\fnsz{-73:55:55.4}&\fnsz{164.0}&\fnsz{10.8}&\fnsz{ 5.2}&\fnsz{ 94.3}&\fnsz{ 5.3}&\fnsz{49.5}\\
\fnsz{ 10}&\fnsz{01:25:30.5}&\fnsz{-72:58:50.8}&\fnsz{147.8}&\fnsz{ 8.2}&\fnsz{ 3.2}&\fnsz{ 57.3}&\fnsz{ 4.2}&\fnsz{48.6}\\
\fnsz{ 11}&\fnsz{01:25:40.1}&\fnsz{-73:32:23.6}&\fnsz{173.4}&\fnsz{12.4}&\fnsz{ 3.7}&\fnsz{ 66.8}&\fnsz{ 3.2}&\fnsz{49.1}\\
\fnsz{ 12}&\fnsz{01:25:40.1}&\fnsz{-74: 3: 3.9}&\fnsz{167.3}&\fnsz{ 4.9}&\fnsz{ 1.8}&\fnsz{ 31.8}&\fnsz{ 3.9}&\fnsz{47.4}\\
\fnsz{ 13}&\fnsz{01:27: 6.6}&\fnsz{-73:28:41.7}&\fnsz{147.0}&\fnsz{ 5.8}&\fnsz{ 2.1}&\fnsz{ 38.5}&\fnsz{ 4.0}&\fnsz{47.8}\\
\fnsz{ 14}&\fnsz{01:27: 8.7}&\fnsz{-73:57: 6.7}&\fnsz{147.0}&\fnsz{ 5.8}&\fnsz{ 2.7}&\fnsz{ 48.3}&\fnsz{ 5.0}&\fnsz{48.1}\\
\fnsz{ 15}&\fnsz{01:27:34.3}&\fnsz{-74: 4:34.1}&\fnsz{147.8}&\fnsz{ 5.0}&\fnsz{ 1.2}&\fnsz{ 21.1}&\fnsz{ 2.6}&\fnsz{46.8}\\
\fnsz{ 16}&\fnsz{01:27:49.0}&\fnsz{-73:17: 3.5}&\fnsz{175.9}&\fnsz{ 5.0}&\fnsz{ 3.5}&\fnsz{ 62.6}&\fnsz{ 7.6}&\fnsz{48.3}\\
\fnsz{ 17}&\fnsz{01:27:54.6}&\fnsz{-73: 4:56.4}&\fnsz{185.1}&\fnsz{ 3.7}&\fnsz{ 1.8}&\fnsz{ 33.4}&\fnsz{ 5.5}&\fnsz{47.2}\\
\fnsz{ 18}&\fnsz{01:28: 3.0}&\fnsz{-74: 5:41.3}&\fnsz{139.6}&\fnsz{ 3.3}&\fnsz{ 1.6}&\fnsz{ 28.2}&\fnsz{ 5.1}&\fnsz{46.9}\\
\fnsz{ 19}&\fnsz{01:28: 8.0}&\fnsz{-71:19:46.1}&\fnsz{204.7}&\fnsz{ 2.5}&\fnsz{ 0.6}&\fnsz{ 11.2}&\fnsz{ 2.7}&\fnsz{45.4}\\
\fnsz{ 20}&\fnsz{01:28:48.4}&\fnsz{-73:46:15.5}&\fnsz{153.5}&\fnsz{ 5.9}&\fnsz{ 3.2}&\fnsz{ 58.9}&\fnsz{ 6.0}&\fnsz{48.3}\\
\fnsz{ 21}&\fnsz{01:29: 8.3}&\fnsz{-73: 6:55.5}&\fnsz{161.8}&\fnsz{15.7}&\fnsz{ 6.9}&\fnsz{125.3}&\fnsz{ 4.8}&\fnsz{50.2}\\
\fnsz{ 22}&\fnsz{01:29:36.7}&\fnsz{-73:41:17.0}&\fnsz{165.1}&\fnsz{ 6.8}&\fnsz{ 2.2}&\fnsz{ 40.8}&\fnsz{ 3.6}&\fnsz{48.0}\\
\fnsz{ 23}&\fnsz{01:29:38.2}&\fnsz{-73: 1:54.7}&\fnsz{184.9}&\fnsz{ 5.8}&\fnsz{ 1.7}&\fnsz{ 31.6}&\fnsz{ 3.3}&\fnsz{47.5}\\
\fnsz{ 24}&\fnsz{01:29:41.0}&\fnsz{-73:17:30.0}&\fnsz{175.0}&\fnsz{ 2.5}&\fnsz{ 3.9}&\fnsz{ 71.6}&\fnsz{17.4}&\fnsz{47.8}\\
\fnsz{ 25}&\fnsz{01:29:50.5}&\fnsz{-73: 3:54.9}&\fnsz{186.7}&\fnsz{ 5.0}&\fnsz{ 1.7}&\fnsz{ 31.7}&\fnsz{ 3.8}&\fnsz{47.4}\\
\fnsz{ 26}&\fnsz{01:29:53.3}&\fnsz{-73:56:56.2}&\fnsz{169.3}&\fnsz{ 6.6}&\fnsz{ 1.2}&\fnsz{ 22.4}&\fnsz{ 2.0}&\fnsz{47.2}\\
\fnsz{ 27}&\fnsz{01:30:14.3}&\fnsz{-73:59:26.6}&\fnsz{159.5}&\fnsz{18.3}&\fnsz{ 5.0}&\fnsz{ 90.1}&\fnsz{ 3.0}&\fnsz{49.9}\\
\fnsz{ 28}&\fnsz{01:30:44.4}&\fnsz{-73:49:42.0}&\fnsz{160.2}&\fnsz{ 2.5}&\fnsz{ 2.5}&\fnsz{ 45.2}&\fnsz{11.0}&\fnsz{47.2}\\
\fnsz{ 29}&\fnsz{01:30:50.3}&\fnsz{-73:42:32.8}&\fnsz{171.5}&\fnsz{ 7.6}&\fnsz{ 0.9}&\fnsz{ 16.7}&\fnsz{ 1.3}&\fnsz{46.9}\\
\fnsz{ 30}&\fnsz{01:31: 0.8}&\fnsz{-73:57:16.1}&\fnsz{150.3}&\fnsz{ 5.8}&\fnsz{ 3.8}&\fnsz{ 68.2}&\fnsz{ 7.1}&\fnsz{48.5}\\
\fnsz{ 31}&\fnsz{01:31:42.3}&\fnsz{-73:46:20.6}&\fnsz{157.2}&\fnsz{ 7.7}&\fnsz{ 3.2}&\fnsz{ 58.7}&\fnsz{ 4.6}&\fnsz{48.6}\\
\fnsz{ 32}&\fnsz{01:31:43.3}&\fnsz{-73:52:25.4}&\fnsz{173.4}&\fnsz{ 5.8}&\fnsz{ 3.4}&\fnsz{ 62.5}&\fnsz{ 6.5}&\fnsz{48.4}\\
\fnsz{ 33}&\fnsz{01:31:56.5}&\fnsz{-73:59:39.0}&\fnsz{177.3}&\fnsz{ 4.7}&\fnsz{ 2.7}&\fnsz{ 49.3}&\fnsz{ 6.3}&\fnsz{47.9}\\
\fnsz{ 34}&\fnsz{01:32: 7.4}&\fnsz{-73:54:43.4}&\fnsz{175.4}&\fnsz{ 7.8}&\fnsz{ 2.2}&\fnsz{ 40.5}&\fnsz{ 3.1}&\fnsz{48.1}\\
\fnsz{ 35}&\fnsz{01:32: 9.9}&\fnsz{-73:30:28.4}&\fnsz{186.1}&\fnsz{ 4.0}&\fnsz{ 1.2}&\fnsz{ 20.9}&\fnsz{ 3.1}&\fnsz{46.6}\\
\fnsz{ 36}&\fnsz{01:32:28.8}&\fnsz{-74: 5: 3.7}&\fnsz{156.9}&\fnsz{ 5.8}&\fnsz{ 1.9}&\fnsz{ 33.8}&\fnsz{ 3.5}&\fnsz{47.6}\\
\fnsz{ 37}&\fnsz{01:32:31.0}&\fnsz{-74: 4: 4.6}&\fnsz{173.0}&\fnsz{ 8.7}&\fnsz{ 2.5}&\fnsz{ 44.6}&\fnsz{ 3.1}&\fnsz{48.3}\\
\fnsz{ 38}&\fnsz{01:32:44.0}&\fnsz{-73:10:36.6}&\fnsz{178.4}&\fnsz{10.7}&\fnsz{ 3.7}&\fnsz{ 67.0}&\fnsz{ 3.8}&\fnsz{49.0}\\
\fnsz{ 39}&\fnsz{01:33:22.8}&\fnsz{-71:48:40.0}&\fnsz{219.2}&\fnsz{ 5.0}&\fnsz{ 3.1}&\fnsz{ 56.3}&\fnsz{ 6.8}&\fnsz{48.1}\\
\fnsz{ 40}&\fnsz{01:33:45.4}&\fnsz{-73:56:49.3}&\fnsz{151.7}&\fnsz{ 5.0}&\fnsz{ 3.9}&\fnsz{ 71.5}&\fnsz{ 8.6}&\fnsz{48.5}\\
\fnsz{ 41}&\fnsz{01:34: 0.8}&\fnsz{-71:56:16.8}&\fnsz{221.2}&\fnsz{ 4.1}&\fnsz{ 2.2}&\fnsz{ 40.3}&\fnsz{ 5.9}&\fnsz{47.5}\\
\fnsz{ 42}&\fnsz{01:34: 1.6}&\fnsz{-71:45:36.7}&\fnsz{217.4}&\fnsz{ 9.4}&\fnsz{ 3.7}&\fnsz{ 67.8}&\fnsz{ 4.3}&\fnsz{48.9}\\
\fnsz{ 43}&\fnsz{01:34:35.3}&\fnsz{-71:59:16.7}&\fnsz{206.5}&\fnsz{16.9}&\fnsz{ 4.5}&\fnsz{ 82.1}&\fnsz{ 2.9}&\fnsz{49.7}\\
\fnsz{ 44}&\fnsz{01:34:40.5}&\fnsz{-72:53:42.3}&\fnsz{208.8}&\fnsz{ 4.9}&\fnsz{ 4.5}&\fnsz{ 81.0}&\fnsz{ 9.8}&\fnsz{48.6}\\
\fnsz{ 45}&\fnsz{01:35:30.8}&\fnsz{-72:56:17.7}&\fnsz{194.0}&\fnsz{ 8.2}&\fnsz{ 4.5}&\fnsz{ 82.4}&\fnsz{ 6.0}&\fnsz{49.1}\\
\fnsz{ 46}&\fnsz{01:37:28.5}&\fnsz{-74:18: 4.7}&\fnsz{151.9}&\fnsz{10.7}&\fnsz{ 3.7}&\fnsz{ 67.2}&\fnsz{ 3.8}&\fnsz{49.0}\\
\fnsz{ 47}&\fnsz{01:41:13.7}&\fnsz{-74:17:12.3}&\fnsz{149.5}&\fnsz{ 6.6}&\fnsz{ 2.3}&\fnsz{ 42.1}&\fnsz{ 3.8}&\fnsz{48.0}\\
\fnsz{ 48}&\fnsz{01:41:34.9}&\fnsz{-73:55:11.1}&\fnsz{166.0}&\fnsz{ 9.9}&\fnsz{ 4.1}&\fnsz{ 74.4}&\fnsz{ 4.5}&\fnsz{49.1}\\
\fnsz{ 49}&\fnsz{01:42:35.9}&\fnsz{-73:49:56.2}&\fnsz{162.7}&\fnsz{ 3.3}&\fnsz{ 1.5}&\fnsz{ 27.1}&\fnsz{ 4.9}&\fnsz{46.8}\\
\fnsz{ 50}&\fnsz{01:43:26.2}&\fnsz{-72:50:40.3}&\fnsz{141.2}&\fnsz{14.8}&\fnsz{ 5.9}&\fnsz{106.5}&\fnsz{ 4.3}&\fnsz{49.9}\\
\fnsz{ 51}&\fnsz{01:44:13.0}&\fnsz{-73:56:33.6}&\fnsz{159.4}&\fnsz{ 6.6}&\fnsz{ 2.4}&\fnsz{ 42.6}&\fnsz{ 3.9}&\fnsz{48.0}\\
\fnsz{ 52}&\fnsz{01:45:22.9}&\fnsz{-74:30: 6.7}&\fnsz{146.3}&\fnsz{ 6.7}&\fnsz{ 1.8}&\fnsz{ 32.8}&\fnsz{ 2.9}&\fnsz{47.7}\\
\fnsz{ 53}&\fnsz{01:45:42.2}&\fnsz{-74:37:44.4}&\fnsz{148.6}&\fnsz{ 4.1}&\fnsz{ 2.0}&\fnsz{ 36.0}&\fnsz{ 5.2}&\fnsz{47.4}\\
\fnsz{ 54}&\fnsz{01:45:53.5}&\fnsz{-73:14:36.7}&\fnsz{177.5}&\fnsz{ 4.9}&\fnsz{ 3.2}&\fnsz{ 58.7}&\fnsz{ 7.1}&\fnsz{48.2}\\
\end{tabular}
\end{center}
\end{table*}
\addtocounter{table}{-1}
\begin{table*}
\begin{center}
\caption{Table of Magellanic Bridge Shell Parameters}
\label{tab:Shelltab}
\vspace{5mm}
\begin{tabular}{lcccccccc}\hline\hline
\fnsz{Shell}&
\fnsz{Right Ascension}& 
\fnsz{Declination}& 
\fnsz{Heliocentric}& 
\fnsz{Exp. Vel.}& 
\fnsz{Radius}&
\fnsz{Radius}&
\fnsz{Dynamic}&
\fnsz{Energy}\\
\fnsz{Number}&
\fnsz{(J2000)}&
\fnsz{(J2000)}&
\fnsz{Vel. (\kms)}&
\fnsz{(\kms)}&
\fnsz{($'$)}&
\fnsz{(pc)}&
\fnsz{Age (Myr)}&
\fnsz{(log erg)}\\
\hline\hline
\fnsz{ 55}&\fnsz{01:46:10.2}&\fnsz{-74:28:26.5}&\fnsz{181.6}&\fnsz{ 5.8}&\fnsz{ 5.5}&\fnsz{ 99.7}&\fnsz{10.4}&\fnsz{49.0}\\
\fnsz{ 56}&\fnsz{01:46:13.9}&\fnsz{-74:38:14.5}&\fnsz{148.6}&\fnsz{ 4.1}&\fnsz{ 1.7}&\fnsz{ 30.9}&\fnsz{ 4.5}&\fnsz{47.2}\\
\fnsz{ 57}&\fnsz{01:46:15.5}&\fnsz{-74:35:28.3}&\fnsz{172.6}&\fnsz{21.4}&\fnsz{ 6.7}&\fnsz{121.6}&\fnsz{ 3.4}&\fnsz{50.4}\\
\fnsz{ 58}&\fnsz{01:47:50.3}&\fnsz{-74: 9: 4.4}&\fnsz{153.0}&\fnsz{ 4.8}&\fnsz{ 2.8}&\fnsz{ 51.1}&\fnsz{ 6.4}&\fnsz{48.0}\\
\fnsz{ 59}&\fnsz{01:48:21.7}&\fnsz{-74:37:16.0}&\fnsz{193.2}&\fnsz{ 2.5}&\fnsz{ 2.0}&\fnsz{ 36.2}&\fnsz{ 8.8}&\fnsz{47.0}\\
\fnsz{ 60}&\fnsz{01:48:50.7}&\fnsz{-74:43:32.9}&\fnsz{156.9}&\fnsz{ 2.5}&\fnsz{ 1.8}&\fnsz{ 33.0}&\fnsz{ 8.0}&\fnsz{46.8}\\
\fnsz{ 61}&\fnsz{01:49:41.1}&\fnsz{-74:21:14.6}&\fnsz{157.7}&\fnsz{11.5}&\fnsz{ 3.6}&\fnsz{ 65.7}&\fnsz{ 3.4}&\fnsz{49.1}\\
\fnsz{ 62}&\fnsz{01:50:47.3}&\fnsz{-74:22: 6.0}&\fnsz{140.4}&\fnsz{ 2.5}&\fnsz{ 1.3}&\fnsz{ 23.4}&\fnsz{ 5.7}&\fnsz{46.4}\\
\fnsz{ 63}&\fnsz{01:50:53.0}&\fnsz{-74:58: 2.2}&\fnsz{194.0}&\fnsz{ 5.0}&\fnsz{ 3.5}&\fnsz{ 62.8}&\fnsz{ 7.6}&\fnsz{48.3}\\
\fnsz{ 64}&\fnsz{01:51:20.8}&\fnsz{-74: 4:17.7}&\fnsz{160.2}&\fnsz{10.7}&\fnsz{ 7.1}&\fnsz{128.1}&\fnsz{ 7.2}&\fnsz{49.9}\\
\fnsz{ 65}&\fnsz{01:51:37.8}&\fnsz{-71:29:54.2}&\fnsz{218.7}&\fnsz{ 3.3}&\fnsz{ 5.0}&\fnsz{ 91.4}&\fnsz{16.6}&\fnsz{48.4}\\
\fnsz{ 66}&\fnsz{01:52: 1.3}&\fnsz{-74: 7:30.4}&\fnsz{193.2}&\fnsz{ 2.5}&\fnsz{ 0.7}&\fnsz{ 12.5}&\fnsz{ 3.0}&\fnsz{45.6}\\
\fnsz{ 67}&\fnsz{01:52:39.9}&\fnsz{-74:28:39.3}&\fnsz{166.7}&\fnsz{ 9.2}&\fnsz{ 3.6}&\fnsz{ 65.4}&\fnsz{ 4.3}&\fnsz{48.9}\\
\fnsz{ 68}&\fnsz{01:52:55.7}&\fnsz{-74:25:55.6}&\fnsz{155.2}&\fnsz{ 2.5}&\fnsz{ 0.9}&\fnsz{ 16.7}&\fnsz{ 4.0}&\fnsz{45.9}\\
\fnsz{ 69}&\fnsz{01:53: 7.1}&\fnsz{-73:59:49.6}&\fnsz{177.5}&\fnsz{16.5}&\fnsz{ 5.4}&\fnsz{ 98.4}&\fnsz{ 3.6}&\fnsz{49.9}\\
\fnsz{ 70}&\fnsz{01:53:42.2}&\fnsz{-71:56:22.3}&\fnsz{221.2}&\fnsz{ 7.4}&\fnsz{ 3.8}&\fnsz{ 68.9}&\fnsz{ 5.6}&\fnsz{48.7}\\
\fnsz{ 71}&\fnsz{01:53:54.5}&\fnsz{-73:55:32.6}&\fnsz{174.5}&\fnsz{16.4}&\fnsz{11.2}&\fnsz{203.9}&\fnsz{ 7.5}&\fnsz{50.8}\\
\fnsz{ 72}&\fnsz{01:53:56.6}&\fnsz{-71:54:25.2}&\fnsz{227.8}&\fnsz{ 4.1}&\fnsz{ 3.5}&\fnsz{ 63.9}&\fnsz{ 9.3}&\fnsz{48.1}\\
\fnsz{ 73}&\fnsz{01:54: 7.5}&\fnsz{-74:39:59.6}&\fnsz{173.2}&\fnsz{14.7}&\fnsz{ 5.7}&\fnsz{104.0}&\fnsz{ 4.2}&\fnsz{49.9}\\
\fnsz{ 74}&\fnsz{01:54:58.2}&\fnsz{-74: 6:18.6}&\fnsz{176.7}&\fnsz{ 7.4}&\fnsz{ 3.5}&\fnsz{ 63.2}&\fnsz{ 5.1}&\fnsz{48.6}\\
\fnsz{ 75}&\fnsz{01:55: 3.3}&\fnsz{-74:18:18.2}&\fnsz{151.1}&\fnsz{ 3.3}&\fnsz{ 5.6}&\fnsz{100.7}&\fnsz{18.3}&\fnsz{48.5}\\
\fnsz{ 76}&\fnsz{01:55:14.3}&\fnsz{-74: 4:22.8}&\fnsz{197.3}&\fnsz{ 3.3}&\fnsz{ 2.2}&\fnsz{ 40.1}&\fnsz{ 7.3}&\fnsz{47.3}\\
\fnsz{ 77}&\fnsz{01:55:26.0}&\fnsz{-71:55:51.5}&\fnsz{231.1}&\fnsz{ 2.5}&\fnsz{ 1.0}&\fnsz{ 18.1}&\fnsz{ 4.4}&\fnsz{46.0}\\
\fnsz{ 78}&\fnsz{01:55:26.6}&\fnsz{-74: 7:25.4}&\fnsz{181.6}&\fnsz{ 2.5}&\fnsz{ 1.4}&\fnsz{ 26.2}&\fnsz{ 6.4}&\fnsz{46.5}\\
\fnsz{ 79}&\fnsz{01:55:52.4}&\fnsz{-74:26:29.3}&\fnsz{165.1}&\fnsz{14.0}&\fnsz{ 3.6}&\fnsz{ 66.1}&\fnsz{ 2.8}&\fnsz{49.2}\\
\fnsz{ 80}&\fnsz{01:56:25.1}&\fnsz{-74: 7:38.5}&\fnsz{154.4}&\fnsz{ 8.2}&\fnsz{ 2.8}&\fnsz{ 50.5}&\fnsz{ 3.7}&\fnsz{48.4}\\
\fnsz{ 81}&\fnsz{01:56:35.2}&\fnsz{-74:24:37.4}&\fnsz{157.8}&\fnsz{ 6.1}&\fnsz{ 4.0}&\fnsz{ 72.2}&\fnsz{ 7.1}&\fnsz{48.6}\\
\fnsz{ 82}&\fnsz{01:57:44.1}&\fnsz{-74:16:46.9}&\fnsz{153.1}&\fnsz{ 4.6}&\fnsz{ 1.7}&\fnsz{ 30.9}&\fnsz{ 4.0}&\fnsz{47.3}\\
\fnsz{ 83}&\fnsz{01:58:17.8}&\fnsz{-74:15: 0.1}&\fnsz{173.4}&\fnsz{15.7}&\fnsz{ 3.7}&\fnsz{ 67.7}&\fnsz{ 2.6}&\fnsz{49.4}\\
\fnsz{ 84}&\fnsz{01:58:25.4}&\fnsz{-74:26:58.8}&\fnsz{150.3}&\fnsz{ 7.4}&\fnsz{ 1.8}&\fnsz{ 32.9}&\fnsz{ 2.7}&\fnsz{47.8}\\
\fnsz{ 85}&\fnsz{02: 0:42.8}&\fnsz{-74:20:22.1}&\fnsz{161.8}&\fnsz{ 4.2}&\fnsz{ 1.8}&\fnsz{ 32.8}&\fnsz{ 4.7}&\fnsz{47.3}\\
\fnsz{ 86}&\fnsz{02: 3:10.0}&\fnsz{-74:47:60.0}&\fnsz{154.3}&\fnsz{ 9.7}&\fnsz{ 4.0}&\fnsz{ 72.4}&\fnsz{ 4.5}&\fnsz{49.0}\\
\fnsz{ 87}&\fnsz{02: 4:56.9}&\fnsz{-74: 9:49.5}&\fnsz{169.3}&\fnsz{16.5}&\fnsz{10.8}&\fnsz{195.8}&\fnsz{ 7.1}&\fnsz{50.8}\\
\fnsz{ 88}&\fnsz{02: 6:37.1}&\fnsz{-74:36:46.6}&\fnsz{175.0}&\fnsz{12.4}&\fnsz{ 4.2}&\fnsz{ 76.0}&\fnsz{ 3.7}&\fnsz{49.3}\\
\fnsz{ 89}&\fnsz{02: 6:40.2}&\fnsz{-74: 3:24.4}&\fnsz{159.2}&\fnsz{ 5.7}&\fnsz{ 4.5}&\fnsz{ 81.6}&\fnsz{ 8.7}&\fnsz{48.7}\\
\fnsz{ 90}&\fnsz{02: 7: 1.1}&\fnsz{-74:16:22.8}&\fnsz{166.8}&\fnsz{ 7.4}&\fnsz{ 4.9}&\fnsz{ 89.4}&\fnsz{ 7.2}&\fnsz{49.1}\\
\fnsz{ 91}&\fnsz{02: 7:14.4}&\fnsz{-74:44:13.9}&\fnsz{190.6}&\fnsz{ 5.4}&\fnsz{ 4.5}&\fnsz{ 81.1}&\fnsz{ 9.1}&\fnsz{48.7}\\
\fnsz{ 92}&\fnsz{02: 7:36.9}&\fnsz{-74:52:55.5}&\fnsz{178.3}&\fnsz{ 2.5}&\fnsz{ 1.4}&\fnsz{ 25.2}&\fnsz{ 6.1}&\fnsz{46.5}\\
\fnsz{ 93}&\fnsz{02: 7:37.6}&\fnsz{-74:31:49.5}&\fnsz{192.3}&\fnsz{ 3.3}&\fnsz{ 2.2}&\fnsz{ 40.8}&\fnsz{ 7.4}&\fnsz{47.4}\\
\fnsz{ 94}&\fnsz{02: 8:22.9}&\fnsz{-74:52:42.2}&\fnsz{191.5}&\fnsz{ 4.1}&\fnsz{ 1.6}&\fnsz{ 28.3}&\fnsz{ 4.1}&\fnsz{47.1}\\
\fnsz{ 95}&\fnsz{02: 9:22.0}&\fnsz{-74:24:50.4}&\fnsz{159.4}&\fnsz{ 3.3}&\fnsz{ 1.9}&\fnsz{ 34.9}&\fnsz{ 6.3}&\fnsz{47.2}\\
\fnsz{ 96}&\fnsz{02: 9:39.6}&\fnsz{-74:52:41.1}&\fnsz{195.6}&\fnsz{ 8.2}&\fnsz{ 1.7}&\fnsz{ 31.5}&\fnsz{ 2.3}&\fnsz{47.8}\\
\fnsz{ 97}&\fnsz{02:10:24.6}&\fnsz{-74:45:41.8}&\fnsz{168.4}&\fnsz{ 4.1}&\fnsz{ 3.0}&\fnsz{ 54.1}&\fnsz{ 7.9}&\fnsz{47.9}\\
\fnsz{ 98}&\fnsz{02:10:43.4}&\fnsz{-75: 7:32.4}&\fnsz{179.3}&\fnsz{ 5.8}&\fnsz{ 3.1}&\fnsz{ 55.5}&\fnsz{ 5.8}&\fnsz{48.2}\\
\fnsz{ 99}&\fnsz{02:10:47.1}&\fnsz{-74: 2:21.4}&\fnsz{169.3}&\fnsz{ 6.6}&\fnsz{ 6.0}&\fnsz{108.7}&\fnsz{ 9.9}&\fnsz{49.2}\\
\fnsz{100}&\fnsz{02:11:55.9}&\fnsz{-74:14:15.8}&\fnsz{161.0}&\fnsz{ 5.0}&\fnsz{ 4.5}&\fnsz{ 81.6}&\fnsz{ 9.9}&\fnsz{48.6}\\
\fnsz{101}&\fnsz{02:11:40.8}&\fnsz{-74:46:36.4}&\fnsz{178.2}&\fnsz{ 9.5}&\fnsz{ 6.0}&\fnsz{108.7}&\fnsz{ 6.8}&\fnsz{49.6}\\
\fnsz{102}&\fnsz{02:12:41.1}&\fnsz{-73:51:45.8}&\fnsz{166.8}&\fnsz{ 2.5}&\fnsz{ 0.7}&\fnsz{ 12.9}&\fnsz{ 3.1}&\fnsz{45.6}\\
\fnsz{103}&\fnsz{02:12:59.9}&\fnsz{-74:50:18.2}&\fnsz{202.2}&\fnsz{ 9.9}&\fnsz{ 2.8}&\fnsz{ 51.6}&\fnsz{ 3.1}&\fnsz{48.6}\\
\fnsz{104}&\fnsz{02:13:33.4}&\fnsz{-74:19:36.4}&\fnsz{180.0}&\fnsz{ 5.8}&\fnsz{ 0.9}&\fnsz{ 16.9}&\fnsz{ 1.8}&\fnsz{46.7}\\
\fnsz{105}&\fnsz{02:13:38.6}&\fnsz{-73:51:40.4}&\fnsz{174.2}&\fnsz{ 9.9}&\fnsz{ 9.1}&\fnsz{165.6}&\fnsz{10.1}&\fnsz{50.1}\\
\fnsz{106}&\fnsz{02:14:16.4}&\fnsz{-74:16: 1.9}&\fnsz{171.8}&\fnsz{ 3.7}&\fnsz{ 2.8}&\fnsz{ 51.2}&\fnsz{ 8.2}&\fnsz{47.8}\\
\fnsz{107}&\fnsz{02:14:19.0}&\fnsz{-74:22:30.8}&\fnsz{162.5}&\fnsz{ 3.6}&\fnsz{ 1.1}&\fnsz{ 20.2}&\fnsz{ 3.4}&\fnsz{46.5}\\
\fnsz{108}&\fnsz{02:14:30.1}&\fnsz{-74:13:30.9}&\fnsz{179.2}&\fnsz{ 6.5}&\fnsz{ 2.7}&\fnsz{ 49.0}&\fnsz{ 4.5}&\fnsz{48.2}\\
\end{tabular}
\end{center}
\end{table*}
\addtocounter{table}{-1}
\begin{table*}
\begin{center}
\caption{Table of Magellanic Bridge Shell Parameters}
\label{tab:Shelltab}
\vspace{5mm}
\begin{tabular}{lcccccccc}\hline\hline
\fnsz{Shell}&
\fnsz{Right Ascension}& 
\fnsz{Declination}& 
\fnsz{Heliocentric}& 
\fnsz{Exp. Vel.}& 
\fnsz{Radius}&
\fnsz{Radius}&
\fnsz{Dynamic}&
\fnsz{Energy}\\
\fnsz{Number}&
\fnsz{(J2000)}&
\fnsz{(J2000)}&
\fnsz{Vel. (\kms)}&
\fnsz{(\kms)}&
\fnsz{($'$)}&
\fnsz{(pc)}&
\fnsz{Age (Myr)}&
\fnsz{(log erg)}\\
\hline\hline
\fnsz{109}&\fnsz{02:14:52.2}&\fnsz{-73:55:31.4}&\fnsz{157.7}&\fnsz{ 4.9}&\fnsz{ 1.4}&\fnsz{ 25.7}&\fnsz{ 3.1}&\fnsz{47.1}\\
\fnsz{110}&\fnsz{02:14:57.8}&\fnsz{-74: 8:59.0}&\fnsz{174.8}&\fnsz{10.5}&\fnsz{ 3.0}&\fnsz{ 54.2}&\fnsz{ 3.1}&\fnsz{48.7}\\
\fnsz{111}&\fnsz{02:15:24.1}&\fnsz{-74: 2:26.0}&\fnsz{176.0}&\fnsz{ 7.9}&\fnsz{ 2.5}&\fnsz{ 45.3}&\fnsz{ 3.4}&\fnsz{48.3}\\
\fnsz{112}&\fnsz{02:16: 0.2}&\fnsz{-74: 1:50.9}&\fnsz{174.2}&\fnsz{ 7.1}&\fnsz{ 2.0}&\fnsz{ 36.1}&\fnsz{ 3.1}&\fnsz{47.9}\\
\fnsz{113}&\fnsz{02:17:31.6}&\fnsz{-74: 9: 5.1}&\fnsz{179.9}&\fnsz{11.6}&\fnsz{ 4.5}&\fnsz{ 81.5}&\fnsz{ 4.2}&\fnsz{49.4}\\
\fnsz{114}&\fnsz{02:19:44.4}&\fnsz{-73:59:40.6}&\fnsz{180.0}&\fnsz{ 7.4}&\fnsz{ 6.8}&\fnsz{123.8}&\fnsz{10.0}&\fnsz{49.5}\\
\fnsz{115}&\fnsz{02:20: 4.0}&\fnsz{-74: 6:35.8}&\fnsz{191.8}&\fnsz{ 6.0}&\fnsz{ 3.7}&\fnsz{ 66.2}&\fnsz{ 6.6}&\fnsz{48.5}\\
\fnsz{116}&\fnsz{02:20:27.5}&\fnsz{-74:18:28.7}&\fnsz{153.0}&\fnsz{ 4.0}&\fnsz{ 1.5}&\fnsz{ 27.3}&\fnsz{ 4.1}&\fnsz{47.0}\\
\fnsz{117}&\fnsz{02:20:37.8}&\fnsz{-74:41:20.6}&\fnsz{163.6}&\fnsz{ 7.4}&\fnsz{ 2.0}&\fnsz{ 36.3}&\fnsz{ 3.0}&\fnsz{47.9}\\
\fnsz{118}&\fnsz{02:21:49.5}&\fnsz{-74:45: 3.8}&\fnsz{168.4}&\fnsz{ 2.5}&\fnsz{ 3.0}&\fnsz{ 54.0}&\fnsz{13.1}&\fnsz{47.5}\\
\fnsz{119}&\fnsz{02:21:58.9}&\fnsz{-74: 4: 8.2}&\fnsz{179.0}&\fnsz{ 6.8}&\fnsz{ 3.5}&\fnsz{ 64.1}&\fnsz{ 5.7}&\fnsz{48.6}\\
\fnsz{120}&\fnsz{02:22:16.1}&\fnsz{-74: 7: 3.6}&\fnsz{187.9}&\fnsz{ 2.8}&\fnsz{ 3.0}&\fnsz{ 54.1}&\fnsz{11.6}&\fnsz{47.6}\\
\fnsz{121}&\fnsz{02:22:31.3}&\fnsz{-74: 7:58.7}&\fnsz{161.8}&\fnsz{ 4.1}&\fnsz{ 1.8}&\fnsz{ 31.8}&\fnsz{ 4.6}&\fnsz{47.2}\\
\fnsz{122}&\fnsz{02:22:38.2}&\fnsz{-74:31:54.4}&\fnsz{174.2}&\fnsz{ 4.9}&\fnsz{ 4.0}&\fnsz{ 72.2}&\fnsz{ 8.7}&\fnsz{48.5}\\
\fnsz{123}&\fnsz{02:23: 0.7}&\fnsz{-74: 7:51.3}&\fnsz{171.7}&\fnsz{ 4.1}&\fnsz{ 2.2}&\fnsz{ 40.0}&\fnsz{ 5.8}&\fnsz{47.5}\\
\fnsz{124}&\fnsz{02:23: 9.2}&\fnsz{-74:17:17.3}&\fnsz{186.5}&\fnsz{10.8}&\fnsz{ 7.0}&\fnsz{126.4}&\fnsz{ 7.0}&\fnsz{49.9}\\
\fnsz{125}&\fnsz{02:23:15.9}&\fnsz{-74: 0:46.8}&\fnsz{186.5}&\fnsz{ 4.2}&\fnsz{ 2.2}&\fnsz{ 40.7}&\fnsz{ 5.8}&\fnsz{47.6}\\
\fnsz{126}&\fnsz{02:24:22.0}&\fnsz{-73: 9:51.4}&\fnsz{210.4}&\fnsz{12.8}&\fnsz{ 8.0}&\fnsz{144.6}&\fnsz{ 6.8}&\fnsz{50.2}\\
\fnsz{127}&\fnsz{02:24:39.6}&\fnsz{-73:13:46.7}&\fnsz{210.9}&\fnsz{15.6}&\fnsz{ 6.0}&\fnsz{108.7}&\fnsz{ 4.2}&\fnsz{50.0}\\
\fnsz{128}&\fnsz{02:25:21.0}&\fnsz{-73:20: 4.4}&\fnsz{195.6}&\fnsz{ 4.9}&\fnsz{ 2.3}&\fnsz{ 42.2}&\fnsz{ 5.1}&\fnsz{47.8}\\
\fnsz{129}&\fnsz{02:25:50.4}&\fnsz{-74: 8:59.0}&\fnsz{176.2}&\fnsz{ 3.6}&\fnsz{ 2.0}&\fnsz{ 37.1}&\fnsz{ 6.2}&\fnsz{47.3}\\
\fnsz{130}&\fnsz{02:25:59.9}&\fnsz{-73:50:55.5}&\fnsz{184.1}&\fnsz{ 6.6}&\fnsz{ 5.8}&\fnsz{104.4}&\fnsz{ 9.5}&\fnsz{49.2}\\
\fnsz{131}&\fnsz{02:27: 8.1}&\fnsz{-74: 0: 2.3}&\fnsz{189.1}&\fnsz{ 5.6}&\fnsz{ 5.3}&\fnsz{ 95.4}&\fnsz{10.2}&\fnsz{48.9}\\
\fnsz{132}&\fnsz{02:28:20.2}&\fnsz{-74: 5: 6.7}&\fnsz{177.5}&\fnsz{ 3.3}&\fnsz{ 2.4}&\fnsz{ 42.7}&\fnsz{ 7.8}&\fnsz{47.4}\\
\fnsz{133}&\fnsz{02:29:27.4}&\fnsz{-74:11:41.0}&\fnsz{172.6}&\fnsz{ 3.3}&\fnsz{ 4.5}&\fnsz{ 81.0}&\fnsz{14.7}&\fnsz{48.2}\\
\fnsz{134}&\fnsz{02:29:35.0}&\fnsz{-73:54:36.6}&\fnsz{175.9}&\fnsz{ 3.3}&\fnsz{ 2.9}&\fnsz{ 53.2}&\fnsz{ 9.7}&\fnsz{47.7}\\
\fnsz{135}&\fnsz{02:30:39.9}&\fnsz{-74: 5:40.3}&\fnsz{192.9}&\fnsz{ 8.2}&\fnsz{ 4.0}&\fnsz{ 72.5}&\fnsz{ 5.3}&\fnsz{48.9}\\
\fnsz{136}&\fnsz{02:30:53.6}&\fnsz{-73:49: 2.4}&\fnsz{180.0}&\fnsz{ 2.5}&\fnsz{ 1.2}&\fnsz{ 22.1}&\fnsz{ 5.4}&\fnsz{46.3}\\
\fnsz{137}&\fnsz{02:32:46.2}&\fnsz{-74: 6:44.8}&\fnsz{193.2}&\fnsz{ 3.3}&\fnsz{ 1.5}&\fnsz{ 27.0}&\fnsz{ 4.9}&\fnsz{46.8}\\
\fnsz{138}&\fnsz{02:33: 8.1}&\fnsz{-73:52: 2.2}&\fnsz{178.3}&\fnsz{ 2.5}&\fnsz{ 2.4}&\fnsz{ 43.3}&\fnsz{10.5}&\fnsz{47.2}\\
\fnsz{139}&\fnsz{02:33:28.3}&\fnsz{-71:14:38.1}&\fnsz{218.7}&\fnsz{ 4.9}&\fnsz{ 6.9}&\fnsz{125.4}&\fnsz{15.3}&\fnsz{49.2}\\
\fnsz{140}&\fnsz{02:34:20.9}&\fnsz{-74:10:31.3}&\fnsz{175.0}&\fnsz{ 2.5}&\fnsz{ 3.5}&\fnsz{ 63.9}&\fnsz{15.5}&\fnsz{47.7}\\
\fnsz{141}&\fnsz{02:36:12.9}&\fnsz{-73:50:28.3}&\fnsz{184.1}&\fnsz{ 3.3}&\fnsz{ 4.0}&\fnsz{ 71.8}&\fnsz{13.1}&\fnsz{48.1}\\
\fnsz{142}&\fnsz{02:36:31.7}&\fnsz{-72:44:43.6}&\fnsz{217.1}&\fnsz{ 8.2}&\fnsz{ 3.6}&\fnsz{ 66.0}&\fnsz{ 4.8}&\fnsz{48.8}\\
\fnsz{143}&\fnsz{02:37:14.5}&\fnsz{-73: 2:31.0}&\fnsz{195.6}&\fnsz{ 3.3}&\fnsz{ 1.1}&\fnsz{ 19.7}&\fnsz{ 3.6}&\fnsz{46.4}\\
\fnsz{144}&\fnsz{02:37:55.5}&\fnsz{-74: 6:38.4}&\fnsz{175.0}&\fnsz{ 2.5}&\fnsz{ 1.0}&\fnsz{ 17.4}&\fnsz{ 4.2}&\fnsz{46.0}\\
\fnsz{145}&\fnsz{02:38:14.9}&\fnsz{-73:45:19.3}&\fnsz{184.0}&\fnsz{ 6.4}&\fnsz{ 3.0}&\fnsz{ 53.8}&\fnsz{ 5.1}&\fnsz{48.3}\\
\fnsz{146}&\fnsz{02:40:44.6}&\fnsz{-74:10: 1.6}&\fnsz{182.4}&\fnsz{ 3.3}&\fnsz{ 3.0}&\fnsz{ 53.7}&\fnsz{ 9.8}&\fnsz{47.7}\\
\fnsz{147}&\fnsz{02:41:57.7}&\fnsz{-72:30: 5.2}&\fnsz{204.7}&\fnsz{ 5.8}&\fnsz{ 4.0}&\fnsz{ 73.3}&\fnsz{ 7.6}&\fnsz{48.6}\\
\fnsz{148}&\fnsz{02:42: 1.7}&\fnsz{-74:11:43.9}&\fnsz{185.7}&\fnsz{ 6.6}&\fnsz{ 2.5}&\fnsz{ 45.0}&\fnsz{ 4.1}&\fnsz{48.1}\\
\fnsz{149}&\fnsz{02:42: 6.9}&\fnsz{-74:14:11.7}&\fnsz{187.3}&\fnsz{ 5.6}&\fnsz{ 2.5}&\fnsz{ 45.4}&\fnsz{ 4.9}&\fnsz{48.0}\\
\fnsz{150}&\fnsz{02:43:10.9}&\fnsz{-74: 3:15.0}&\fnsz{194.0}&\fnsz{ 6.6}&\fnsz{ 2.7}&\fnsz{ 49.5}&\fnsz{ 4.5}&\fnsz{48.2}\\
\fnsz{151}&\fnsz{02:43:22.1}&\fnsz{-74: 4:18.7}&\fnsz{180.4}&\fnsz{ 2.2}&\fnsz{ 1.2}&\fnsz{ 22.6}&\fnsz{ 6.1}&\fnsz{46.2}\\
\fnsz{152}&\fnsz{02:43:58.2}&\fnsz{-74:17: 2.4}&\fnsz{185.7}&\fnsz{ 3.3}&\fnsz{ 1.2}&\fnsz{ 22.5}&\fnsz{ 4.1}&\fnsz{46.6}\\
\fnsz{153}&\fnsz{02:44:21.8}&\fnsz{-73:24:49.1}&\fnsz{191.5}&\fnsz{ 4.1}&\fnsz{ 3.3}&\fnsz{ 59.0}&\fnsz{ 8.6}&\fnsz{48.0}\\
\fnsz{154}&\fnsz{02:44:25.6}&\fnsz{-75:12: 3.8}&\fnsz{201.9}&\fnsz{ 3.9}&\fnsz{ 2.5}&\fnsz{ 45.3}&\fnsz{ 6.9}&\fnsz{47.6}\\
\fnsz{155}&\fnsz{02:44:31.8}&\fnsz{-73:19:53.9}&\fnsz{193.2}&\fnsz{ 4.1}&\fnsz{ 3.5}&\fnsz{ 63.4}&\fnsz{ 9.2}&\fnsz{48.1}\\
\fnsz{156}&\fnsz{02:46: 1.2}&\fnsz{-73:59:27.8}&\fnsz{191.5}&\fnsz{ 4.1}&\fnsz{ 1.1}&\fnsz{ 20.0}&\fnsz{ 2.9}&\fnsz{46.6}\\
\fnsz{157}&\fnsz{02:46: 7.5}&\fnsz{-74: 8:29.2}&\fnsz{187.4}&\fnsz{ 4.9}&\fnsz{ 2.4}&\fnsz{ 44.0}&\fnsz{ 5.3}&\fnsz{47.8}\\
\fnsz{158}&\fnsz{02:46:50.2}&\fnsz{-73:30:29.9}&\fnsz{193.2}&\fnsz{ 4.1}&\fnsz{ 1.3}&\fnsz{ 23.3}&\fnsz{ 3.4}&\fnsz{46.8}\\
\fnsz{159}&\fnsz{02:47:46.1}&\fnsz{-74: 0:12.2}&\fnsz{198.1}&\fnsz{ 5.8}&\fnsz{ 2.5}&\fnsz{ 44.6}&\fnsz{ 4.6}&\fnsz{48.0}\\
\fnsz{160}&\fnsz{02:48:22.5}&\fnsz{-75:49:42.1}&\fnsz{201.4}&\fnsz{ 4.1}&\fnsz{ 2.4}&\fnsz{ 44.1}&\fnsz{ 6.4}&\fnsz{47.7}\\
\fnsz{161}&\fnsz{02:50:26.4}&\fnsz{-75:52:17.8}&\fnsz{200.6}&\fnsz{ 3.3}&\fnsz{ 4.2}&\fnsz{ 75.9}&\fnsz{13.8}&\fnsz{48.2}\\
\fnsz{162}&\fnsz{02:53:55.7}&\fnsz{-74: 5:22.7}&\fnsz{184.1}&\fnsz{ 3.3}&\fnsz{ 3.9}&\fnsz{ 70.9}&\fnsz{12.9}&\fnsz{48.1}\\
\fnsz{163}&\fnsz{02:54: 3.4}&\fnsz{-73:39:47.1}&\fnsz{210.2}&\fnsz{ 2.7}&\fnsz{ 2.0}&\fnsz{ 36.3}&\fnsz{ 8.2}&\fnsz{47.0}\\
\end{tabular}
\end{center}
\end{table*}

Magellanic Bridge shell parameters are collated and graphically
represented in plots against Right Ascension in Figs.
\ref{fig:ra-mb}a-\ref{fig:ra-mb}d. A comparison of Magellanic Bridge
and SMC shell histogram parameters is shown in Table~\ref{tab:Comptab}
and plotted against Right ascension in
Figs~\ref{fig:ra_smc-mb}a-\ref{fig:ra_smc-mb}c.  The shell parameters
are shown as log histograms in Figs \ref{fig:histo}a-\ref{fig:histo}d.

\begin{table*}
\begin{center}
\begin{tabular}{rcc}\hline\hline
  &\small{Magellanic Bridge}& \small{Small Magellanic Cloud} \\ \hline
  \small{Mean Shell Radius,(R$_s$)} &58.6pc&91.9 pc \\ 
  \small{$\sigma$(R$_{s}$)} &33.2 pc&65.5 pc\\ \small{Mean Expansion
    Velocity, (V$_s$)}&6.5 \kms\&10.3 \kms\ \\ 
  \small{$\sigma$(V$_{s}$)}&3.8 \kms\&6.3 \kms\ \\ \small{Mean Dynamic
    Age, (T$_s$)} &6.2 Myr&5.7 Myr \\ \small{$\sigma$(T$_{s}$)}&3.4
  Myr&2.8 Myr\\ \small{Mean Energy, (L$_s$)}& 48.1 log(erg)& 51.8
  log(erg)\\ \hline\hline
\end{tabular}
\caption{Comparison of properties of Magellanic Bridge Shells (this study) and
  Small Magellanic Cloud Shells (Staveley-Smith et al. 1997). The mean
  and standard deviation of each property are given}
\label{tab:Comptab}
\end{center}
\end{table*}

\subsubsection{Statistical analysis}

Table~\ref{tab:Comptab} shows that the average shell radius and
expansion velocity of the Magellanic Bridge shell population are
smaller than for the SMC population, while the average kinematic age is
slightly larger.  The dispersions of shell radius and expansion
velocity of the Bridge population are also slightly lower than the SMC
population, while the dispersion for kinematic age, which is a
dependent of both the shell radius and expansion velocity, is slightly
larger for the Bridge population. We also see that the mean energy of
the shell population is considerably lower in the Magellanic Bridge
than for the SMC. To some extent, the disagreement between the mean
energy and mean radii here is probably an effect of a different shell
selection criteria used for this survey (see also
Fig.\ref{fig:ra_smc-mb}).  This is discussed in more detail in
Section~\ref{sec:limits}.

\subsubsection{Right Ascension plots}

\begin{figure*}

  \centerline{
    \psfig{file=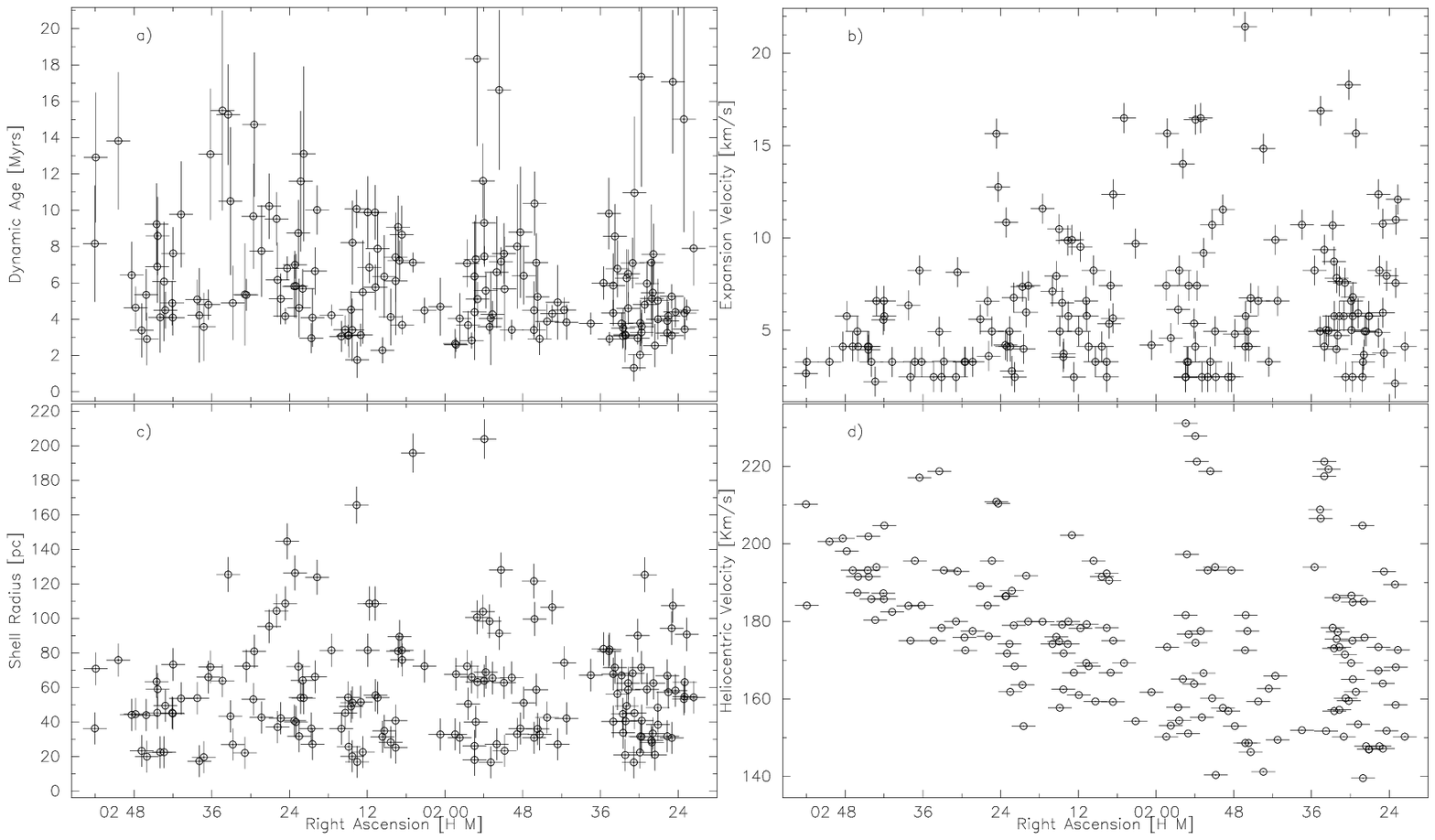}}
\caption{Parameters of Magellanic Bridge shells,  plotted against
  RA. \textit{a)Dynamic Age}, \textit{b)Expansion Velocity},
  \textit{c)Shell radius}, \textit{d)Heliocentric Velocity}}
\label{fig:ra-mb}
\end{figure*}

Figs \ref{fig:ra-mb}a -\ref{fig:ra-mb}d show the Dynamic age,
expansion velocity, shell radius and heliocentric velocity for each of
the shells plotted against Right Ascension. A number of observations
can be made immediately from these Right Ascension plots:
\begin{itemize}
\item{There does not appear to be a gradient of shell age with RA, as
    shown in Fig.\ref{fig:ra-mb}a, although there is a subtle
    tendency of older shells to be found at higher RA.  This is
    discussed further in Section~\ref{sec:limits}.}
\item{The mean expansion velocity appears to decrease towards the LMC
    (see also Fig.\ref{fig:ra-aves}). This figure shows also that the
    dispersion of expansion velocity is reduced after 2\hr20\min,
    where the values become less scattered.}
\item{There is no general trend of shell radius with RA, although once
    again, there appears to a subtle departure into larger radii
    shells at higher RAs.  This is also discussed in
    Section~\ref{sec:limits}.}
\item{The RA-Vel plot (Fig.\ref{fig:ra-mb}d highlights the smoothly
    increasing Heliocentric velocity of the \hi\ shells of the
    Magellanic Bridge towards the LMC, and shows a few shells arranged
    in apparent loops and filaments.}
\end{itemize}

Fig.\ref{fig:ra-aves} shows the variation of the mean dynamic age,
mean expansion velocity and mean shell radius against RA along the
sampled region in the Magellanic Bridge.  These plots include
parameters of shells found during this survey only, and are averaged
in five bins across the observed region.  This plot highlights some of
the above trends.

The mean expansion velocity decreases with RA, slowly at low RA, then
more quickly at higher RA $>$\sm 2\hr12\min.  We see that the mean
shell radius appears to be increasing with RA, peaking at \sm
2\hr24\min, before rapidly reducing to low radii.  The mean dynamic
age is relatively low for regions less than RA\sm2\hr24\min, after
which we see a dramatic increase by \sm45 per cent at higher RA.  Since
dynamic age is proportional to radius, and inversely proportional to
expansion velocity, the peak in mean radius at \sm 2\hr24\min, and the
low expansion velocity is manifested as a higher mean dynamic age at
this RA.  Although the mean radius decreases drastically at higher RA,
the expansion velocity has become low enough to allow the dynamic age
to remain high.  A closer look at Fig.\ref{fig:ra-mb}a shows that, in
fact, older shells also exist close to the SMC. These shells are not
representative of the region and are washed out in the binning
process.  The observation that the {\em mean} dynamic age is larger at
higher RA is still true.

\begin{figure}
  \centerline{\psfig{file=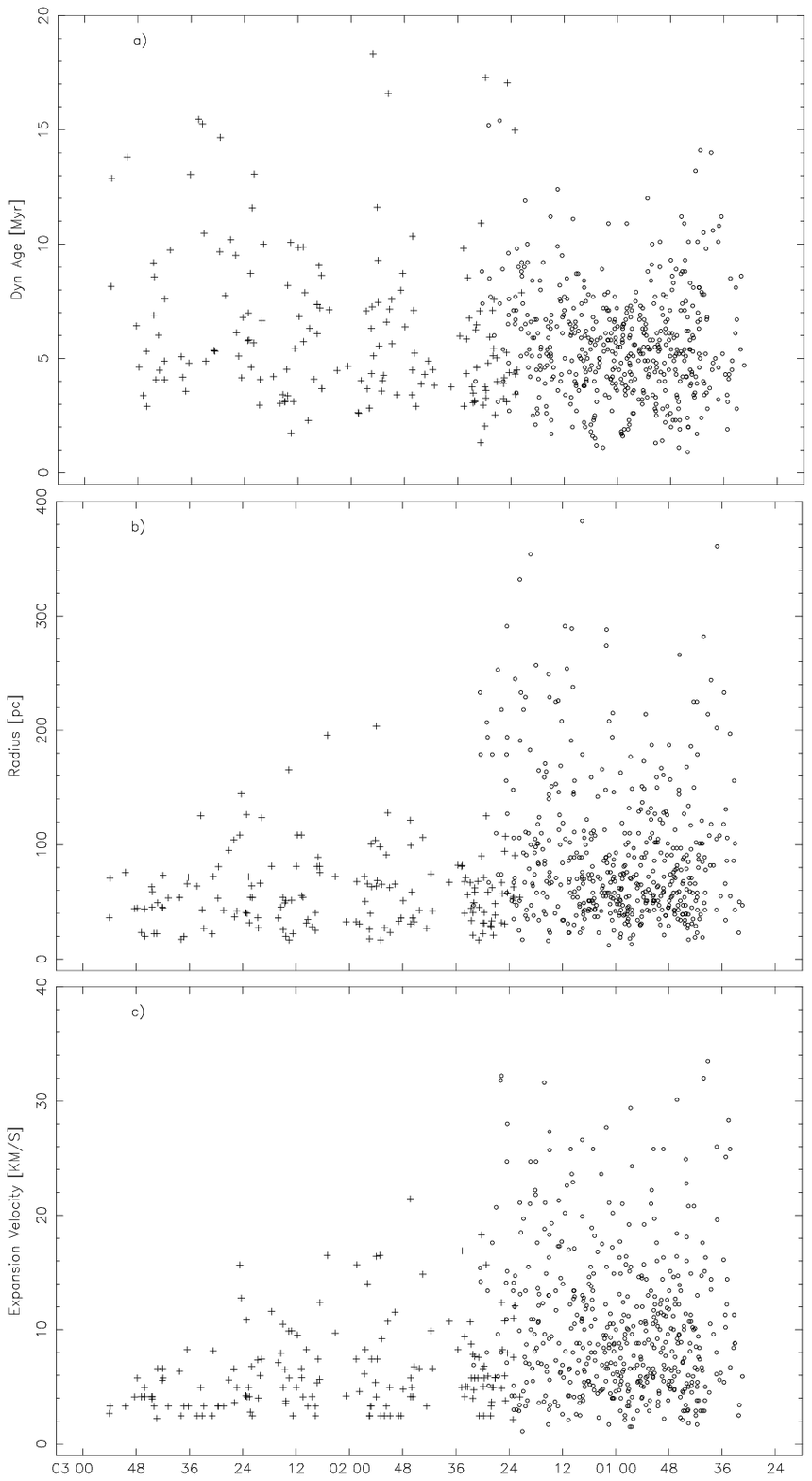}}
\caption{Shell properties of the SMC and Magellanic Bridge as a function of Right
  Ascension.  Top to bottom: a)Dynamic age, b)Shell radius c)Expansion
  Velocity.  Crosses represent shells in the Magellanic Bridge (This
  survey), while small circles represent data from the SMC shell
  survey (Staveley-Smith et al, 1997).}
\label{fig:ra_smc-mb}
\end{figure}

Fig. \ref{fig:ra_smc-mb}a-\ref{fig:ra_smc-mb}c compares the dynamic
age , the radii and the expansion velocities of the Magellanic Shells
with those of SMC shells (Staveley-Smith et al, 1997).  We see from
these figures that there is no obvious discontinuity in the mean of
the shell kinematic age between the SMC and the Magellanic Bridge.
This suggests that there is a continued flow of matter between the two
systems. The figures showing the shell radii and expansion velocities
(Fig.\ref{fig:ra_smc-mb}c) along the Bridge and SMC reveal a sharp
discontinuity corresponding to the position of overlap between the two
surveys.  These figures highlight differences is shell selection
criteria, discussed further in Section~\ref{sec:limits}.

\begin{figure}
  \centerline{\psfig{file=fig8.ps,width=8cm,angle=-90}}
\caption{Mean Shell dynamic Age
  {\it (solid line, axis on the left side)}, Expansion velocity {\it
    (dash line, left axis)}, and Shell radius {\it (dot-dash line,
    right axis)}, errorbars mark the standard error of the mean.}
\label{fig:ra-aves}
\end{figure}

\subsubsection{Histogram Analysis}\label{sec:histo}
\begin{figure*}
  \centerline{
    \psfig{file=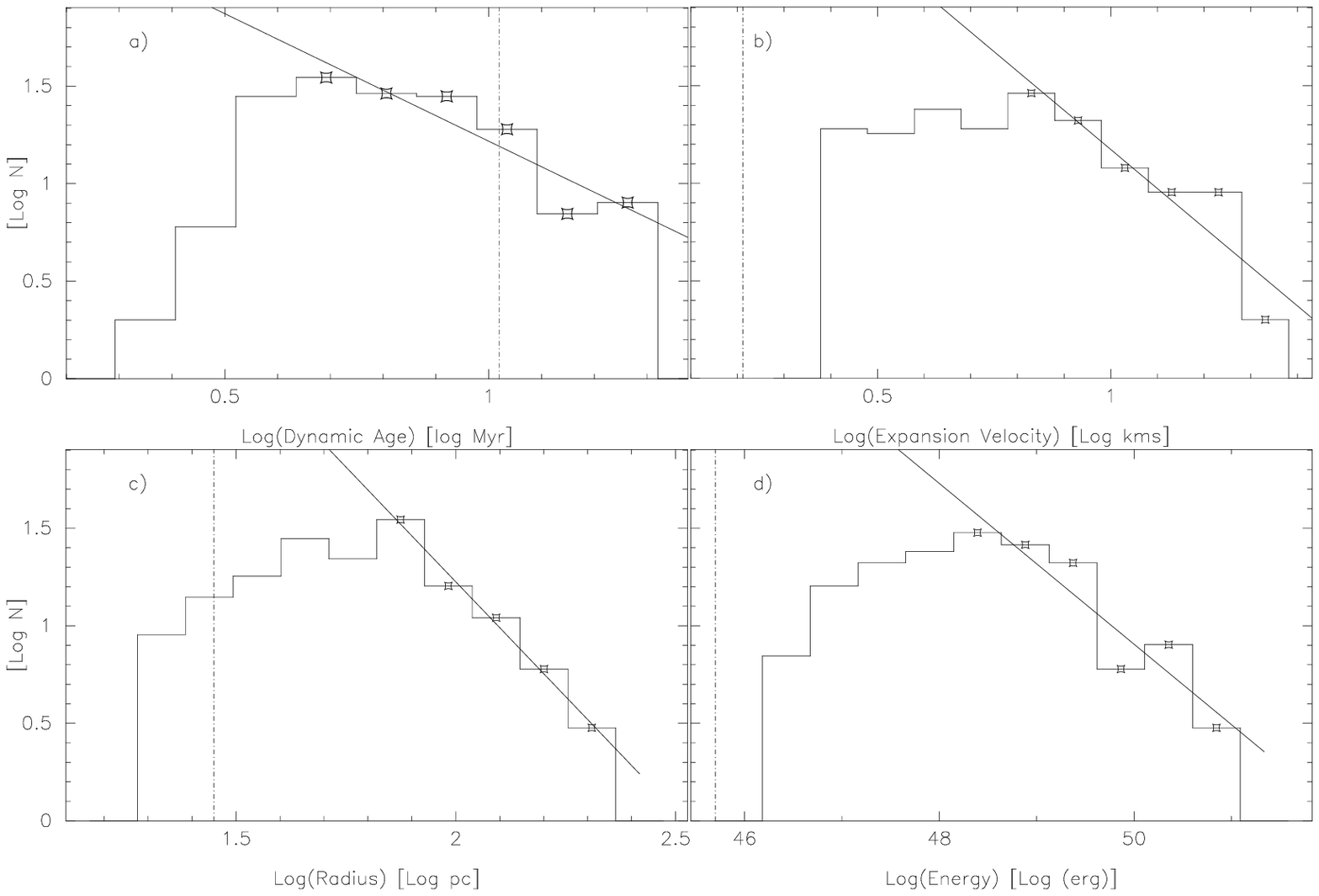}
    }
\caption{Number frequency of Magellanic Bridge Shell parameters:
  \textit{a)Dynamic age, b) Expansion Velocity, c) Radius} and
  \textit{d)Energy}. The vertical \textit{dot-dash} line marks the
  limits imposed by angular resolution (98\asec) and velocity
  resolution (1.61\kms). As the distributions of Shell parameters
  follow a logarithmic law, we are able to determine a characteristic
  slope in log (N)- log space.}
\label{fig:histo}
\end{figure*}

Histograms of various properties of the Magellanic Bridge shells are
shown in Fig.\ref{fig:histo}. Shell parameters follow a
logarithmic distribution (eg. Oey \& Clarke, 1997), and the frequency
histograms in log space can be fit with linear model.
Power law slopes are fitted to each of
the parameters: dynamic age, expansion velocity and radius, and are
compared with those from the SMC and \hoii\ in
Table~\ref{tab:powerlawcomp}.

\begin{table}
\begin{center}
\begin{tabular}{lccc}
  \hline\hline &\small{Holmberg II}&\small{SMC}&\small{Bridge}\\\hline
  \small{Number of shells}&\small{51}&\small{509}&\small{163}\\ 
  \small{Expansion Velocity,
    $\alpha_v$}&\small{2.9$\pm$0.6}&\small{2.8$\pm$0.4}&\small{2.6
    $\pm$0.6}\\ \small{Shell Radius,
    $\alpha_r$}&\small{2.0$\pm$0.2}&\small{2.2$\pm$0.3}&\small{3.6$\pm$0.4}\\\hline\hline
\end{tabular}
\caption{Power law of shell radius and expansion velocity
  ($\alpha_r$ and $\alpha_v$) for Holmberg II (Puche, 1992), the SMC
  (Stanimirovi\'c, 1999) and the Magellanic Bridge.  The Slope in linear
  space ($\alpha$) is related to the slope in log space ($\gamma$) by
  $\alpha=1-\gamma$}
\label{tab:powerlawcomp}
\end{center}
\end{table}

These values for each of the slopes vary slightly, depending on the
bin size used for each histogram.  The tabulated
figures represent the average $\alpha$ (where $\alpha$ is the slope in
linear space and is related to the slope in log space $\gamma$ with
$\alpha=1-\gamma$), while the errors represent the range of $\alpha$
while varying the number of bins from 10 to 20.

It can be seen that the power-law fit of the expansion velocity for
shells populating the Magellanic Bridge appears to be reasonably
consistent with that of the Holmberg II shell population and with the
SMC population.  Although the slope of the fit to shell radius
distribution is considerably steeper, this is most probably due to an
deficiency of larger diameter shells (and is discussed in
Section~\ref{sec:limits}).  When comparing these three systems, we
should bear in mind that the kinematic conditions of the \hoii\ galaxy
are not necessarily reproduced in the SMC and the Magellanic Bridge.
Specifically, \hoii\ is a disk galaxy and is not obviously tidally
perturbed.

\section{Distribution of Blue associations}\label{sec:corr}
\subsection{Spatial correlation with \hi\ expanding shells}

Fig.\ref{fig:shellolay} shows an integrated intensity map of the
Magellanic Bridge overlaid with the positions and sizes of \hi\ shells,
as well as the OB associations found within the Bridge.  The latter
was initially compiled by Batinelli \& Demers (1992) and extended to
cover the rest of the Bridge and the SMC by Bica \& Schmitt (1995).
The limiting magnitude of this association survey is V=20.0, however,
in an attempt to eliminate bright foreground stars, only associations
where (B-V)$>$0.0 were included . A visual examination of
Fig.\ref{fig:shellolay} shows a general correlation between \hi\
column density, the number density of expanding shells, and with the
number density of OB associations.  The detailed alignment of the
associations with shell centres is very poor however, although
grouping of young blue clusters about higher \hi\ density regions can
be seen in many instances, and in particular about the rims of some
larger shells and filaments.  A more quantitative study of the
relative distributions of OB associations and \hi\ expanding shells
shows that \sm40 per cent of Magellanic Bridge shells have one or more
associations within a distance equal to its radius.  If we assume for
a moment that associations are responsible for generating the shell, a
displacement over the mean shell radius (60 pc), in over a time
interval equal to the mean kinematic age (6 Myr) would require a
velocity of only \sm1kms\ (assuming no inclination of the
trajectory of the association to the plane of the sky).  However, any motion of
the OB association relative to shell would result in the shell having
a significantly deformed and non-spherical shape, and would therefore
have been excluded from the survey.  The poor spatial correlation
statistic of OB associations and \hi\ expanding shells is contrary to
the popular theory of the formation of stellar wind-driven \hi\
expanding shells (Weaver et al, 1977).  Similarly however, we should
also bear in mind that the mean Magellanic Bridge shell age is
approximately equivalent to that of an O type star, and any related
stellar association may be too faint to have been included in the OB
catalogue.  A study of the spatial correlation of the SNe and \hi\
shell population of the \hoii\ galaxy has been conducted by Rhode et
al. (1999).  This study was designed to test the hypothesis of stellar
wind and SN explosions acting as the engine for the expansion of a \hi\
shells.  The conclusions from this analysis were that under the
assumption of a normal initial mass function, the OB cluster
brightness was such that the \hi\ shell distribution was strongly
inconsistent with the theory of formation by SNe.

\subsection{Properties of nearby \hi}
\begin{figure*}
  \centerline{ \psfig{file=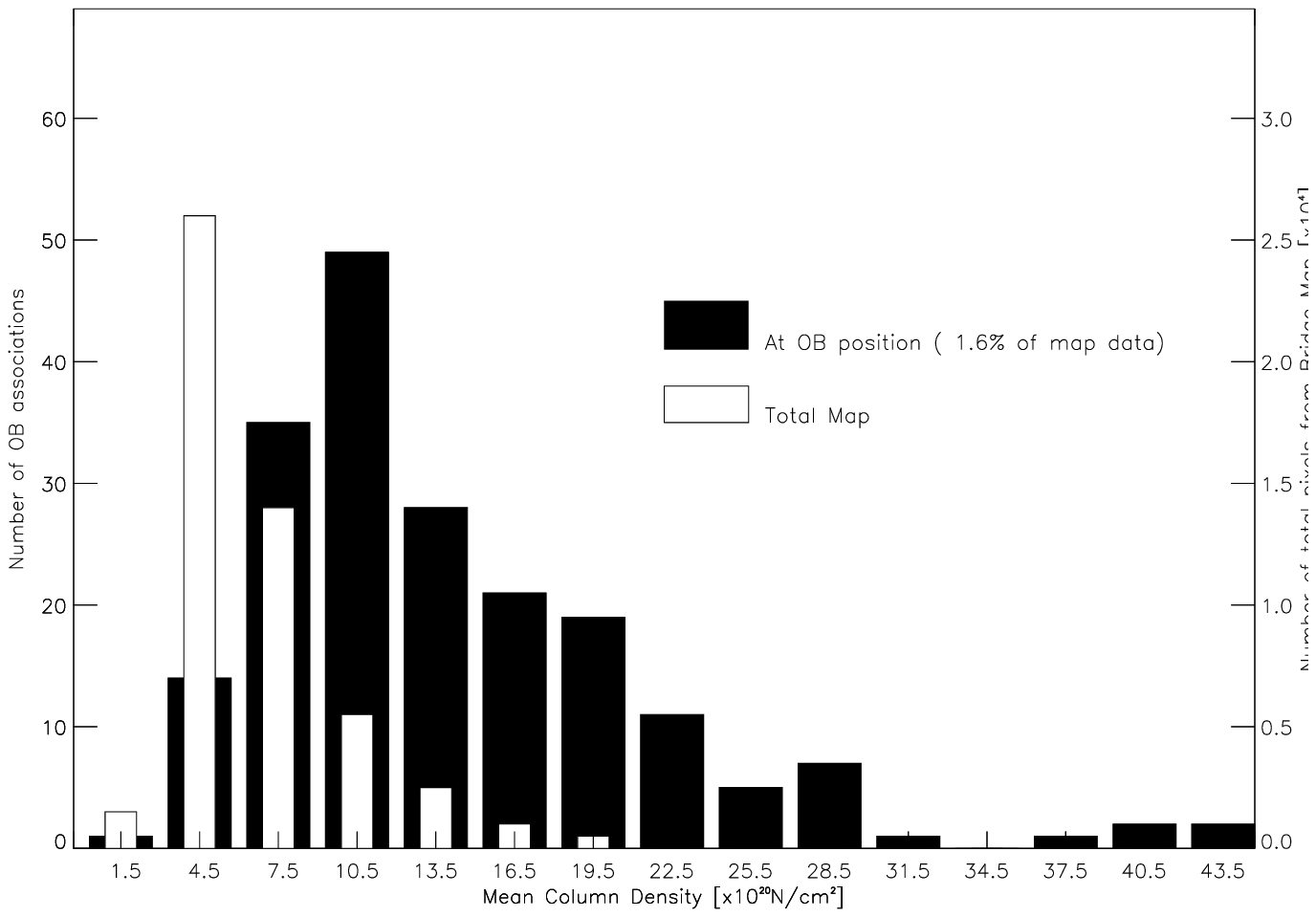}}
\caption{{\it Black and left hand axis:} 
  Histogram of mean column density within a 90\asec$\times$90\asec
  square centred around each OB association (from catalogue by Bica \&
  Schmitt 1995).  {\it White and right hand axis:} Histogram of mean
  column density of entire map.  Map is binned into
  90\asec$\times$90\asec resolution.}
\label{fig:assoccorr}
\end{figure*}

To quantitatively test the spatial correlation of OB associations and
\hi\ column density in the Magellanic Bridge, the mean column density
of a 90\asec (3 pixel) box centred on each of the catalogued OB
association positions is presented as a histogram in
Fig.\ref{fig:assoccorr} (black columns).  Overlaid on this is a
second histogram, representing the entire map itself (white columns).
The histogram shows that \sm50 per cent of the catalogued OB association
positions correlate with a mean column density of
$\gesim$1.2$\times$10$^{21}$cm$^{-2}$, only 8 per cent correspond to column
densities equal to or less than half that density and \sm10 per cent are
associated with regions of column density greater than
2.4$\times$10$^{21}$cm$^{-2}$.  We find that these are similar to
results by Demers \& Grondin, who found that stellar positions
correlate with column densities \sm10$^{21}$cm$^{-2}$, and that very
few associations can be found to correlate with low \hi\ column
densities.

The different distributions of the two histograms in
Fig.\ref{fig:assoccorr} confirm that the mean intensities around the
positions of the associations are a unique subset of the total
dataset, and not simply a random sample, although we should bear in
mind that by selecting a 90\asec$\times$90\asec box at each
association position, we sample less than 2 per cent of the entire map area.

\begin{figure*}
  \centerline{ \psfig{file=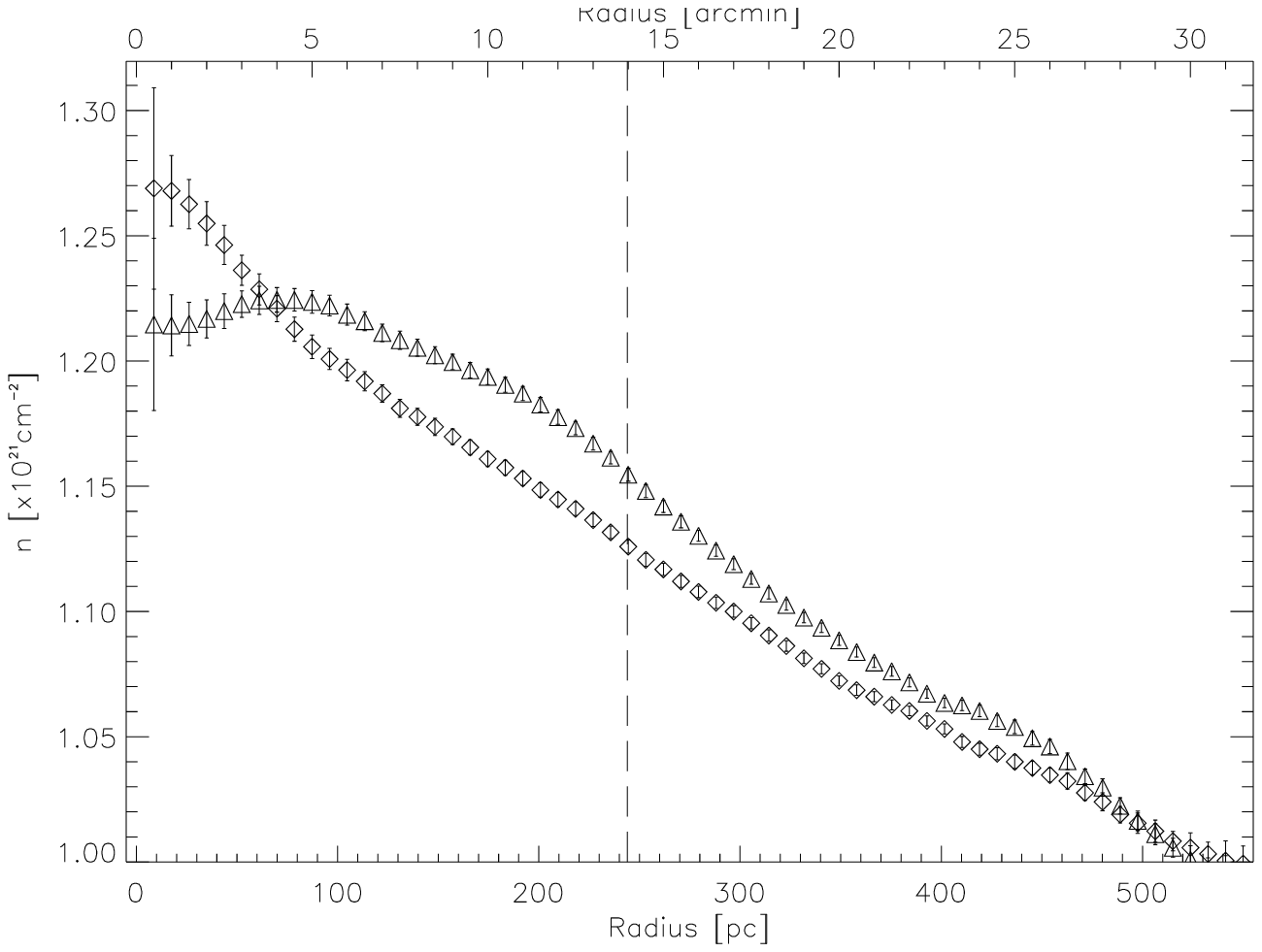}}
\caption{Diamonds show mean \hi\ vs radius centred on OB
  positions, while triangles show mean \hi\ as a function of radius,
  offset from OB centres by \sm5\amin southward.  A positive departure
  is apparent for radii $<$80pc (\sm5\amin).  The offset highlights
  the fact that the OB associations are generally associated with a
  local peak in integrated \hi.  Vertical dashed line shows limit of
  spatial resolution from Matthewson \& Ford (1984) dataset. Error
  bars mark one standard error.}
\label{fig:hi_vs_rad}
\end{figure*}

Fig.\ref{fig:hi_vs_rad} shows the mean column density variation as a
function of distance from the centres of each catalogued OB
association.  A linear fit, with a slope of \sm-0.5$\times$10$^{18}$
cm$^{-2}$pc$^{-1}$, represents the general decline of column density
away from the main \hi\ filaments where OB associations are found.
However, there is a significant departure at short radii.  This
appears to be due to an excess of \hi\ within \sm80pc of the OB
associations, and is in the opposite sense to that found by Grondin \&
Demers (1993) for OB stars.

Data used by Grondin \& Demers was that obtained from the Parkes
telescope by Mathewson \& Ford (1984), and has a resolution of
\sm14\amin (\sm244pc), whereas the minimum resolution of the dataset
used here is \sm98\asec (\sm 28.5pc). Fig.\ref{fig:hi_vs_rad} also
shows the mean integrated \hi\ as a function of radius, offset 5\min
(\sm90pc) south for each OB association.  This line shows a peak in
\hi excess at a distance equivalent to the offset, showing that the
excess is real, and is centred on the positions of the OB associations.

\section{\ha\ regions}\label{sec:ha}
\begin{figure}
  {\psfig{file=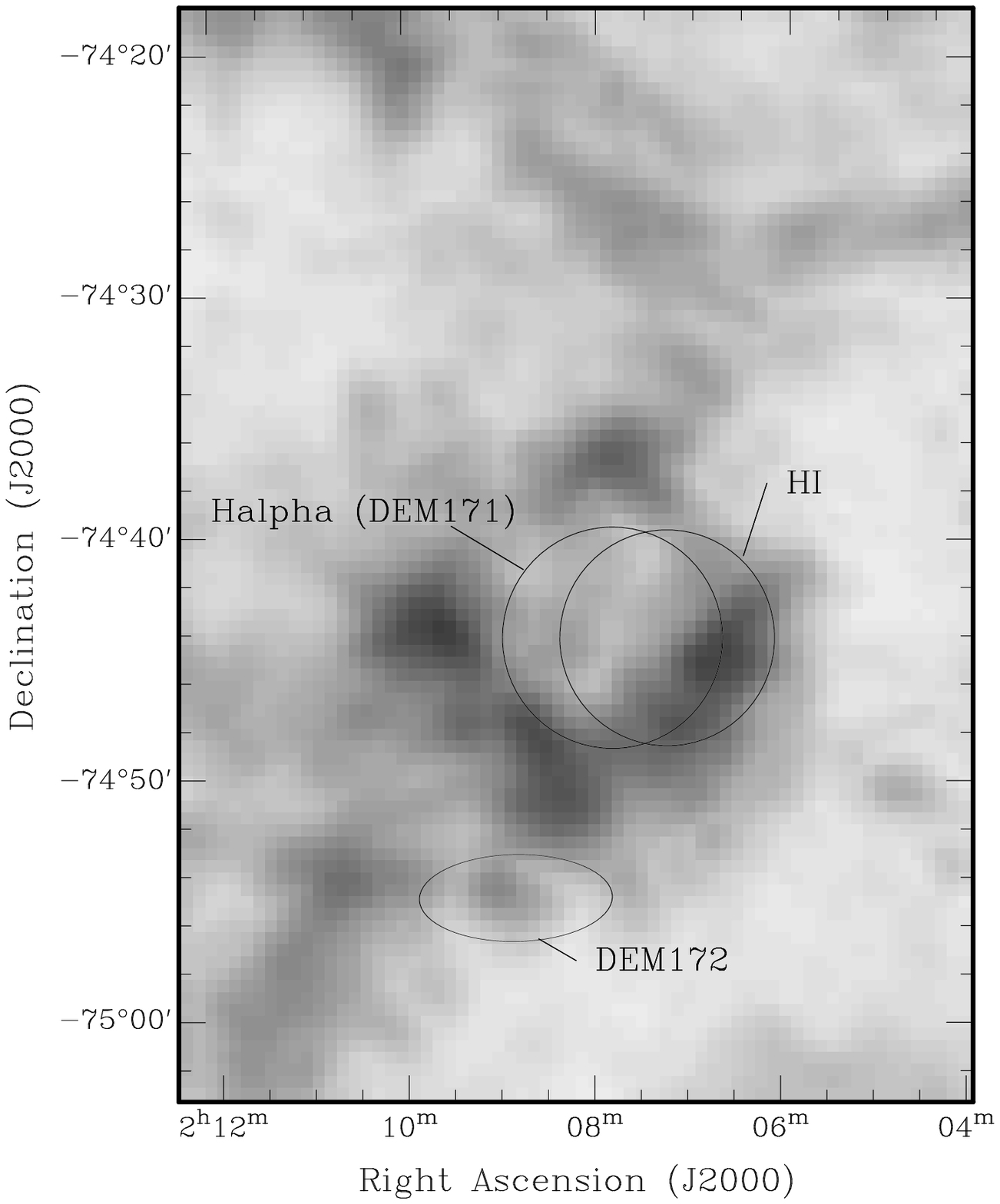,width=9cm}}
\caption{\hi\ (shell \#91) and numbered \ha\ 'nebulous' regions (from Meaburn
  1986), overlaid on magnified Ra-Dec peak pixel map.  The \hi\ shell
  is shown to be closely aligned with the \ha\ shell (labelled \ha).
  The feature to the south, labelled \#172, shows the position of one
  of the \ha\ 'nebulous regions' corresponding to a small local \hi\
  peak intensity maximum.  The reason for the \hi\ and \ha\ shells
  offset can be seen in Fig.\ref{fig:ha-spec}, where the actual
  expanding shell appears to have a higher positive central velocity
  than the ring, and is at a slightly different RA. The figure has a
  linear transfer greyscale ranging from \sm8 to 73\sm Kelvin}
\label{fig:ha}
\end{figure}

A large shell within the Bridge at \mbox{RA 02\hr07\min14\sec,}
\mbox{Dec $-$74\deg44\amin14\asec} (J2000) was detected in \ha\ and
measured by Meaburn (1986), Parker (1998) and by Graham et al. (2001).
This \ha\ shell, labelled DEM171 by Meaburn (1986), is found to be
closely aligned to an identifiable expanding \hi\ shell at
\sm RA 02\hr08\min8\sec, Dec $-$74\deg42\amin46\asec.  This shell is found to be
a typical example of the Magellanic Bridge shell population, and is
parameterised and listed as \#91 in Table~\ref{tab:Shelltab}.  \ha\
parameters as estimated by Meaburn (1986), Parker (1998) and by Graham
(2001) are compared in Table~\ref{tab:hacomp}.  A map of the \hi\ peak
brightness temperature, with the positions of the \ha\ shell as
measured by Meaburn, and the \hi\ shell \#91 (this paper), is shown in
Fig.\ref{fig:ha}.

\begin{table*}
\begin{center}
\begin{tabular}{lcccc}\hline\hline
  &\small{Meaburn (\ha)}&\small{Parker (\ha)}&\small{Graham
    (\ha)}&\small{Shell 91 (\hi)}\\\hline \small{Right
    Ascension(J2000)}&\small{02\hr07\min50\sec}&\small{02\hr07\min56\sec}&\small{02\hr07\min14\sec}&\small{02\hr07\min14\sec}\\ 
  \small{Declination (J2000)}&\small{$-$74\deg
    44\amin14\asec}&\small{$-$74\deg44\amin06\asec}&\small{$-$74\deg44\amin14\asec}&\small{$-$74\deg44\amin14\asec}\\ 
  \small{Radius}&\small{3.93'}&\small{4.3'}& \small{\sm4'}
  &\small{4.5$\pm$0.5'}\\ \small{Apparent age}&\small{5 Myr
    (wind)}&\small{--}&\small{0.53 Myr}&\small{9$\pm$2Myr}\\ 
  \small&\small{8 -210Myr (SN)}&&&\\ \small{Heliocentric
    Vel}&\small{--}&\small{--}&\small{192.5\kms}&\small{190$\pm$2\kms}\\ 
  \small{Expansion velocity}&&&\small{37.0\kms}&\small{5$\pm$2\kms}\\ 
  \small{Suggested Source}&\small{O star stellar wind, SN}&\small{PN,
    SNR, WR shell}&\small{WR}&\small{--}\\ \hline\hline
\end{tabular}
\end{center}
\caption{Comparison of parameters of the \hi\/\ha\ Shell DEM171 (\hi\
  shell\#91) as observed in \hi\ and \ha\ (Meaburn 1986, Parker 1998,
  Graham 2001), and in \hi\ (this paper)}
\label{tab:hacomp}
\end{table*}

The ionising mechanism of this \hi/\ha\ shell is still unknown. Meaburn
(1986) suggests a single O star is responsible for the illumination of
the \ha\ shell. Parker has commented that UV source FAUST 392
corresponds closely to the centre of the shell, is probably a low
surface brightness Planetary nebula (PN) or even a supernova remnant,
although the low energy derived in the present study does not suggest
a typical supernova as the mechanism for this shell.  Graham et
al. (2001) have located a Wolf-Rayet (WR) candidate within the shell
rim, and suggest that this object may be responsible for causing the
expansion of the shell.  A number of OB associations are distributed
around the high density \hi\ rim of this region (Bica \& Schmitt 1995),
and at present, this is the only \hi\ shell within the Magellanic
Bridge that can be unmistakably attributed to a stellar origin.

\begin{figure}
  \centerline{\psfig{file=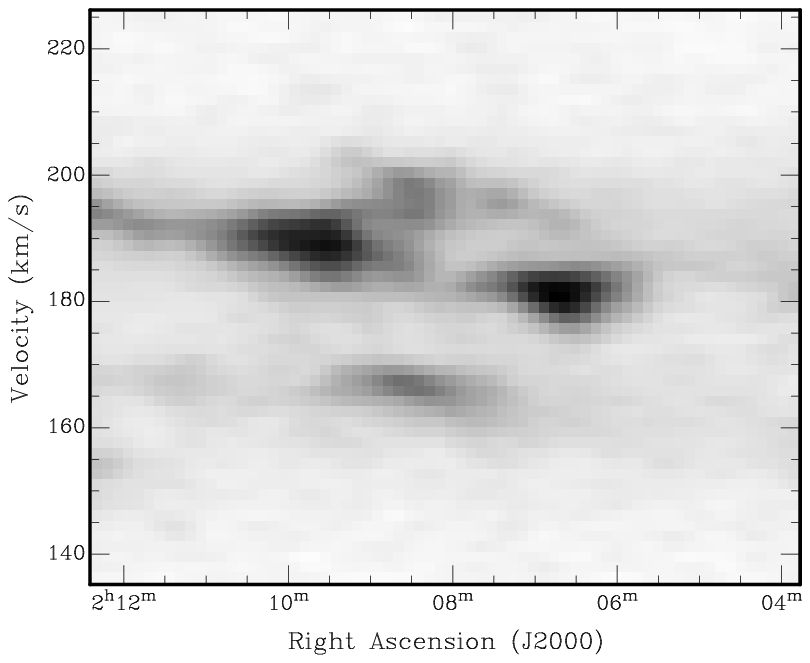}}
\caption{RA-Velocity slice at Dec=\sm$-$74\deg44\amin34\asec, centred on
  DEM171/\hi\ shell \#91.  The temperature range is 70 K to $-$0.4K,
  with a linear transfer function. This shows the receding and
  approaching sides of the expanding shell at \sm185\kms\ and
  \sm196\kms.  We see here that the actual centre of the shell appears
  to be shifted slightly to lower RA, and does not correspond to the
  centre of the \hi\ ring in the RA-Dec projection from
  Fig.\ref{fig:ha}}
\label{fig:ha-spec}
\end{figure}

Table~\ref{tab:hacomp} shows that the age and radius determined here
for shell\# 91 as derived from the \hi\ data is in general agreement
with these parameters determined from \ha\ data by Meaburn (1986),
although there is considerable discrepancy of the expansion velocity,
and hence the kinematic age with findings of Graham et al. (2001).
Fig.\ref{fig:ha-spec} shows the velocity slice at Declination
\mbox{\sm$-$74\deg44\amin34\asec}. The two peaks, at Heliocentric
velocities of \sm185\kms\sm196\kms\ are those corresponding to the
approaching and receding sides of the shell.  We do not find any \hi\
emission peaks corresponding to those in \ha\ as measured by Graham et
al.

In addition, another \ha\ region (region \#172) parameterised by
Meaburn (1986), is found to correlate well with a similarly shaped
high density \hi\ region, as shown in Fig.\ref{fig:ha}.

\section{Factors affecting the Survey}\label{sec:limits}
This survey has selected a sample of shells that, compared with the
\hi\ shell population of the SMC, appears to be relatively deficient
in large-radii shells (See Fig.\ref{fig:ra_smc-mb}b). This is almost
certainly due to a different and tighter selection function. This
survey demands a regular and identifiable ring shape in all three
projections before such a structure can be accepted as an expanding
\hi\ shell.  Incomplete, or significantly distorted shells cannot be
accurately parameterised, and it is not always clear that such
structures are genuine expanding \hi\ shells.  Other \hi\ shell
surveys have used a more relaxed criteria and have permitted partially
incomplete ring shapes to be classified as an expanding shell. Given
that this survey is sensitive to the same range of scales of the SMC
surveys made by Staveley-Smith et al. (1997) and Stanimirovi\'c et al.
(1999), this then leads to question the apparent tendency for large
shells to be more susceptible to fragmentation and distortion.  McCray
\& Kafatos (1987), Ehlerova et al. (1997) and others have determined
that for thin walled expanding shells, instabilities will cause the
shell to fragment after some time.  These authors calculate that
shells will tend to self-destruct from intrinsic instabilities at
radii \sm1 kpc, which is much larger than the maximum radii of the
Magellanic Bridge shell population.  There must therefore be
additional processes catalysing the fragmentation, or otherwise
affecting the integrity of the large-scale \hi shells within this population.

\subsection{Deformation by secondary starformation}
Star formation occurring within the compressed gas comprising the
shell wall may be responsible for deformation of the shell shape.
Mass and energy loss from a star forming within the compressed gas of
an existing shell may blow open a secondary wind shell in the primary
shell wall, leading to deformation of the primary shell, and
ultimately, a departure from the signature shell shape. A shell
expanding into a low ambient density region will not accumulate a high
density rim as quickly as one embedded in a higher density.  As star
formation usually occurs only after a threshold column density is
reached, we might not expect shells that are expanding into a low
ambient density medium to be as readily disturbed by secondary star
formation.  Secondary star formation within the shell wall has been
observed in the SMC (Stanimirovi\'c, 1999), while small \hi\ shells
clustered within the wall of a larger \hi\ shell have also been
observed in the LMC (Kim et al.  1999).

\subsection{Deformation from density stratification}
Shell-like features, such as blow-outs, or chimneys, that were not
included in this catalogue, were found occasionally throughout the
cube.  A blow-out, or a tunnel can develop by an expanding shell
forming close to a region of much lower relative density. The
expanding gas can blow through the boundary separating the two
densities, such as through the wall of a gas cloud, and into the low
density region.  Such structures can also form through the merging of
two expanding shells, and have been observed in the Galaxy (eg
McClure-Griffiths et al, 2000), as well as other galaxies (eg.  Ott et
al., 2001). Under these conditions, the calculation of the dynamic
age, which is based on an assumption of constant and homogeneous
ambient gas density, is incorrect. A study of these shell-like
structures will be included in a future project.

\subsection{Size limitations}
Any constraints on shell radii imposed by the extent of the gas in the
Magellanic Bridge are not considered to be significant: the height of
the high \hi\ density region in the Bridge, in Declination, is almost
four times the diameter of the largest shell found from this survey,
although it is of the same order of the diameter of the largest
supershell in in the SMC (Stanimirovi\'c, 1999). The largest shell
radius found during this survey was 11.7\amin, equivalent to \sm204
pc, and the radius of the largest supershell found in the SMC was
\sm910 pc.

\subsection{Tidal stretching}
Given the mechanism of formation of the Bridge, one possible mechanism
of deformation is tidal stretching. However, of the shells surveyed,
there does not appear to be any significant tendency for elongation or
stretching along the SMC-LMC direction, implying that tidal shearing
of the shells is not a significant cause of distortion over the time
scales considered here.

\subsection{Change of environment in Magellanic Bridge}
Vishniac (1983) has suggested that local
inhomogeneities in the ambient \hi\ gas may cause the shell to distort
from perfect spherical symmetry and we note that Fig.\ref{fig:ppix1}
shows the \hi\ has a complex and turbulent structure down to 98\asec
(\sm29pc). The chaotic nature of the gas comprising the Magellanic
Bridge may be responsible for premature fragmentation of the \hi\ 
shells. Shells fragmented in this way do not satisfy criteria {\em i}
and are not included in this catalogue.
We notice that this survey has uncovered a region within the
Magellanic Bridge containing a slight excess of older shells
Fig.\ref{fig:ra-aves}).  These shells have significantly slower
expansion velocity, and although they also have a slightly smaller
radius, the net result is a mean dynamical age which is a factor of
two or so greater than the rest of the Magellanic Bridge population.
The transition of these two regions appears to be at around RA
2\hr15\amin, with the region containing the older shells lying to
the east of this.
\begin{figure*}
  \centerline{\psfig{file=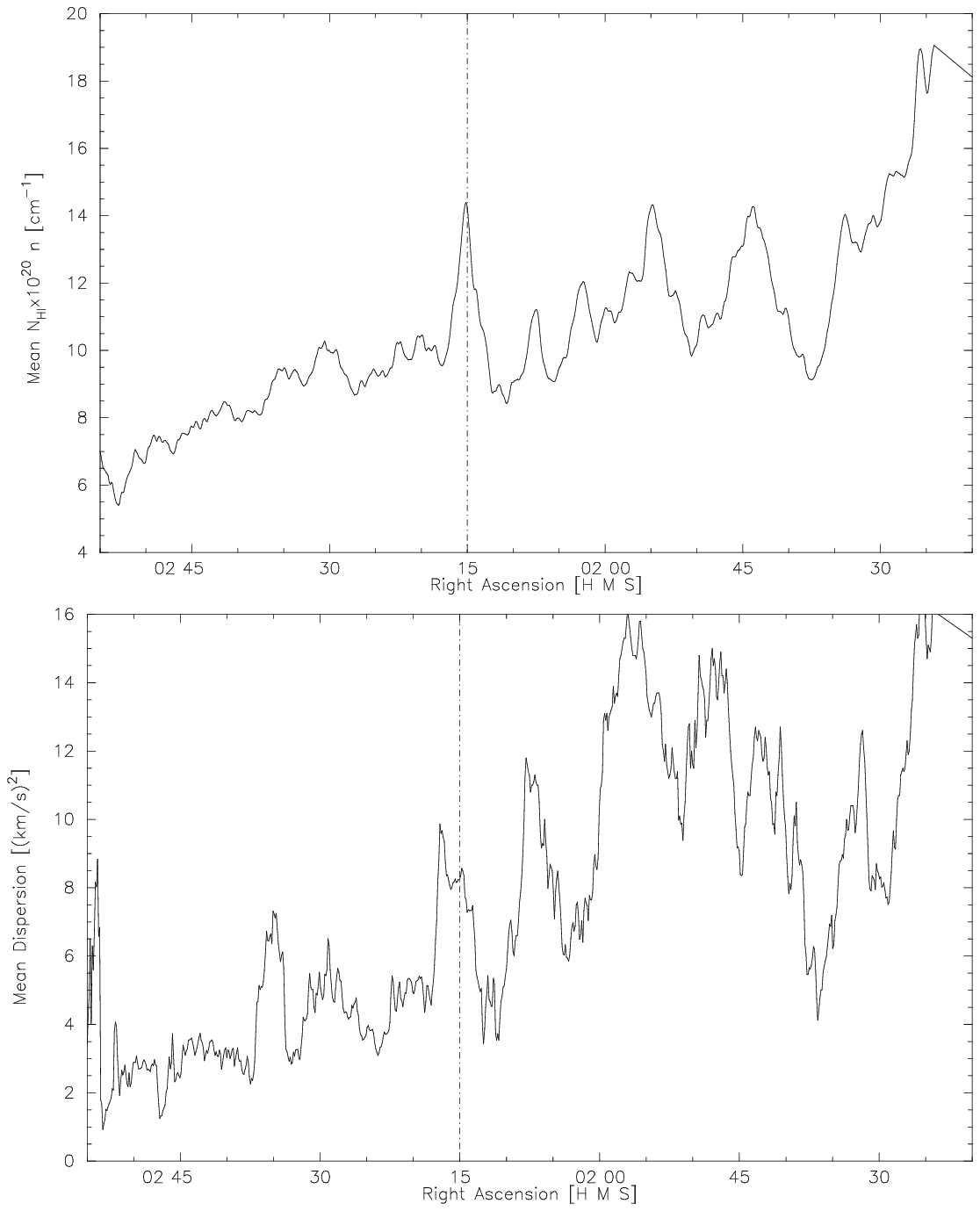}}
\caption{The variation of \hi\ integrated intensity (a) \textit{(top)} and \hi\
  velocity dispersion (b) \textit{(bottom)} with Right Ascension. The
  data is averaged over the high \hi\ column density region across a
  selected area of Fig.\ref{fig:ppix1} (approx $-$73\deg40\amin to
  $-$74\deg40\amin).  These plots suggest sudden change of the
  properties of \hi\ at Right Ascensions higher than \sm2\hr
  15\min.}
\label{fig:inthi-ra}  
\end{figure*}

The average integrated \hi, and \hi\ velocity dispersion for the
central region containing the higher \hi\ column density are plotted in
Figs~\ref{fig:inthi-ra}a and \ref{fig:inthi-ra}b. It can be seen that the
fluctuation of the mean column density is relatively low at RA higher
than 2\hr15\min.  Similarly, the velocity dispersion becomes
somewhat lower above this RA, although there are still some larger
scale variations present, as well as a decreasing gradient with RA. In
general however, the velocity dispersion and \hi\ column density are
smaller and with less variation above 2\hr15\min, suggesting that
this region is less dynamic than the western part. A reference to
Fig.\ref{fig:ppix1}a shows that the large loop mentioned in section
\ref{sec:cuberesults} re-joins the Bridge at approximately this right
Ascension. Also, the velocity bimodality in Fig~\ref{fig:ppix1}b
appears to terminate at this same RA.  We suggest that the younger \hi\
expanding shells west of 2\hr15\min are more quickly ruptured from
the relatively higher turbulence, and possibly from secondary star
formation, while those populating the eastern side remain intact for a
longer time, possibly because of the more quiescent nature of the
ambient gas.  It is unclear at this point what relationship the large
loop might have with the change of the \hi\ environment.

\section{Discussion of the Stellar wind model.}\label{sec:discussion}
\subsection{Energy deposition}
The associations catalogued by Batinelli \& Demers (1992), which were
later included in the catalogue by Bica et al. (1995), constitute many
poorly populated (mean N\sm8) associations and clusters. Bica E.
(priv.  communication) mentions that the associations and clusters
become more populated towards the SMC, and that a few of the
associations may be composed of later type stars, although the
majority are of O-B type.  We can see that the shells so far uncovered
are of very low energy.  Given that the mean shell energy is rather small
when compared with the standard approximation of the energy for a
single O5 type star (\sm10$^{51}$ erg Lozinskaya, 1992), poorly
populated associations, comprising low-mass, later type stars, might
be capable of producing low energy structures and may be responsible
in some of these cases found in the Bridge.  It is curious however,
that we do not observe a larger fraction of expanding shells centred
about many of these observable associations and clusters, and that the
regions surrounding the associations are not depleted in \hi.  The
lack of spatial correlation of \hi\ shells and an obvious energetic
source has been noted before however (e.g. Rhode et al. 1999), and
alternative scenarios to the stellar wind engine are discussed in
Section~\ref{sec:alternatives}.

A total of 198 associations and clusters are found within the observed
Bridge region where 163 shells are found.  The SMC survey constitutes
\sm500 shells, and includes 987 associations and clusters from the
same catalogue.  We find that we have almost one shell per association
or cluster in the Bridge, whereas we have almost two associations or
clusters for each shell within the SMC.  This could reflect different
formation and destruction mechanisms in the two regions.  For example,
the less turbulent nature of the environment of the Bridge may allow
shells to develop for a longer time.

The relatively higher shell population in both cases also calls into
question the idea of stellar wind-driven expanding shells, although
given a similar kinematic age of shells in the Bridge and the SMC, the
relatively higher ratio of shells per association in the Bridge might
indicate a relatively less turbulent environment in this region
(following the discussion from above).

We find that the mean energy of Bridge \hi\ shells is a relatively small
value of \sm1.9$\times$10$^{49}$ ergs, and a total energy of
\sm3.2$\times$10$^{51}$ erg. Table\ref{tab:Shelltab} shows that only
six of the 163 shells have an energy greater than 10$^{50}$ ergs, and
none are greater than 10$^{51}$ ergs (for an ambient density of
\sm0.06 cm$^{-2}$).

The total \hi\ mass of the observed Magellanic Bridge region was
calculated in Section~\ref{sec:cuberesults} to be 1.5$\times$10$^{8}$
M\sun, giving a shell power per \hi\ mass of 2.1$\times$10$^{43}$
ergs/M\sun.  Using a mass of 3.8 $\times$10$^{8}$ M\sun\ for the SMC,
and a total shell energy of \sm6.7$\times$10$^{54}$ erg
(Staveley-Smith et al, 1997 and Stanimirovi\'c et al, 1999), we
calculate a power per \hi\ mass of 1.8$\times$10$^{46}$ ergs/M\sun\ for
the SMC. We see that the shells populating the Bridge are
significantly less powerful per mass than those of the SMC. We see
also that the median SMC shell energy is \sm10$^{50.2}$ erg, and that
the median Magellanic Bridge shell energy is some orders lower at
\sm10$^{48}$ erg.  However, given the strict shell selection criteria,
these relative values should be considered as lower limit only.

\subsection{The Ages of Bridge shells, and Bridge OB associations}
It can be seen that the average dynamic age of \sm6.2 Myr
(Table~\ref{tab:Comptab}) of the Magellanic Bridge shell population is
far younger than the date of the most recent Magellanic Cloud
interaction, shown by simulations to be approximately 200 Myr ago (eg
Gardiner, Sawa \& Fujimoto, 1994).  There is also a lack of agreement
with ages of OB associations, determined through isochrone fitting by
Grondin, Demers \& Kunkel (1992) and Battinelli \& Demers (1998), to
be in the range 10-25 Myr.  These values differ with the average shell
dynamic ages calculated here by up to a factor of almost four. A
possible source of error in the determination of the dynamic age is
suggested by Shull \& Saken (1995) who observe that the fraction of
Wolf-Rayet stars in an OB association peaks for associations of around
\sm 5 Myr old. The extra luminosity input from the WR stars can
accelerate the radii and expansion velocity of the stellar wind shell,
and result in a mis-estimation of the dynamic age by up to a factor of
three.  The average dynamic age calculated here is very close to this
time of peak Wolf-Rayet population, although even allowing for
re-acceleration by WR stars does still not bring the dynamic age to
one that is comparable with the association ages estimated by Grondin,
Demers \& Kunkel (1992) and Demers \& Battinelli (1998).  It is
difficult, therefore, to become convinced that the majority of the
spherical expanding \hi\ structures in the Magellanic Bridge are
powered by stellar winds.

Figs.\ref{fig:ra_smc-mb}a-c show that, despite the apparent lack of large
radius shells in the Magellanic Bridge shell sample, the kinematic age
distribution is consistent with that of the SMC population.  This may
suggest that the environmental conditions for the evolution of \hi\
shells is similar in the two systems.

During a survey of SMC \hi\ shells, Staveley-Smith et al. (1997) also
found a large number of shells with ages that were too young to be
conveniently related to a period of star formation triggered by the
Magellanic cloud encounter. Under the assumption that \hi\ shells are
commonly driven by stellar winds, they suggest that as the Magellanic
system is close to perigalacticon, it may be experiencing sufficient
tidal disruption to stimulate a more recent period of star formation.

It is worth noting the assumptions made in calculating the kinematic
age are not necessarily satisfied in every study. The equations used
in this paper, as derived by Weaver et al. (1977), assume a
homogeneous medium, and relate only to the expansion phase of the
shell.  The Magellanic Bridge can be seen to have structure and
dynamics down to the smallest observed scales, and it is difficult to
asses the extent to which the inhomogeneities affect the estimations
of kinematic age and shell luminosity. The kinematic ages estimated in
this paper are used more as means to compare between systems where
data have been subjected to the same assumptions, rather than an
measure of the absolute age of the shell. Alternative models exist,
and give rise to different estimates of the shell parameters. For
example, a supernova model has been derived by Chevalier (1974).
Using this model,the mean Magellanic Bridge shell energy is estimated
to be \sm5$\times$10$^{49}$, which is an order of magnitude larger
than the estimates made using Weavers stellar wind model.  The Weaver
model has been used in studies of the SMC and \hoii, and to preserve
some continuity, we have used it also in this study.

\section{Alternatives to the stellar wind model}\label{sec:alternatives}
The lack of spatial correlation of the stellar population and the
Shell population imply that other mechanisms may be operating to
produce the observed spherical, expanding \hi\ structures.  These
include energy depositions through Gamma Ray Bursts (GRBs), High
velocity Cloud impacts (HVCs) and ram pressure effects.
\subsection{Gamma Ray Bursts}
The origins of GRBs are not well understood, however, common models
include the collisions of Neutron stars (e.g. Blinnikov et al. 1984) or
the collapse of super-massive objects (Paczy\'nski, 1998).  These
events can deposit \sm10$^{52}$ ergs into the ISM (Wijers et al.,
1998), \sm10 per cent of which is imparted as kinetic energy into the local
medium. Measurements of the gamma ray flux by Wijers et al. (1998),
have led to an estimate of the probability of a GRB event of once per
40 million years per galaxy.  These estimates assume that the GRB
events release energy isotropically and without beaming, although
beamed GRB events have been modelled to produce spherical structures
after \sm5 Myr (Ayal \& Piran 2001). If apply this probability
estimate applies to the Magellanic Bridge, and assume the age of the
Bridge is \sm200 Myr, as found by computer simulations (e.g. Gardiner,
Sawa \& Fujimoto 1994), we can estimate \sm5 Gamma ray Burst events
since its creation.  The number of shells found in the Bridge is
clearly in excess of this, and there is no plausible way to justify
the source of the majority of \hi\ expanding features as GRB relics.
However, caution should be taken when assuming the Magellanic Bridge
will have the same kind of GRB frequency as a galaxy.  The Bridge has
a relatively small total mass, and it is likely to have a
significantly lower Star formation rate. This, however, can only
decrease the estimated GRB frequency.  Additionally, the shell
energies, as derived using both Weaver and Chevalier models, are three
and two orders of magnitude respectively lower than the expected
energy from a GRB. The low GRB frequency, and the insufficient shell
energy suggest that GRB events are not a dominant mechanism for shell
formation and expansion within the Magellanic Bridge.

\subsection{HVCs}
Simulations of HVC impacts into a galactic disk have been generated by
Tenerio-Tagle et al (1986).  Eight different scenarios were
investigated, where the density, radius and velocity of the impacting
HVC were varied.  In general, it was found that HVCs of densities \sm
0.1 - 1 cm$^{-3}$, impacting at 300 \kms\ were found to generate a
spherical expanding void, with radii \sm35-95 pc after \sm5 Myr, and
energies of \sm10$^{48-49}$.  More dense clouds, moving at higher
velocities were capable of penetrating deeper into the gas disc and
creating cylindrical holes of radii \sm70pc in diameter after \sm5 Myr
and depositing \sm10$^{49-51}$ ergs into the ISM.  From
Table~\ref{tab:Comptab}, we see that the mean energy of the Magellanic
Bridge shell population is more compatible with the former scenarios,
however it should be borne in mind that this estimation of the shell
energy is made assuming a stellar wind model, although certainly the
radii of these simulations are consistent with those measured from the
Magellanic Bridge shell population.  It should also be noted that
these simulated HVCs were impacting into a medium of density 1
cm$^{-3}$. The estimates of the Magellanic Bridge, from
Section~\ref{sec:cuberesults}, are two orders of magnitude lower than
this.  We might therefore expect that an HVC cloud impact, such as the
ones simulated by Tenorio-Tagle et al., would have a more destructive
effect than the simulations show. Such destructive impacts may not
generate a complete, spherically expanding shell which would
consequently be excluded from this survey.  As such, although we can
see from Fig.\ref{fig:shellolay} that the distribution of the surveyed
\hi\ expanding shells appears to be cheifly confined to regions of
higher \hi, we cannot confidently rule out this mechanism of shell
formation for the entire Bridge shell population on the basis of this
non-uniform distribution.

\subsection{Ram Pressure}
Bureau \& Carignan (2002) have suggested that the stripping action of
the intergalactic medium on infalling cluster members could also
generate holes and tunnels.  The feasibility of this hypothesis has
been studied through simulations only tentatively, although with
promising results.  This model suffers from similar setbacks as the
infalling HVC model in the case of the Magellanic Bridge: We would
expect the distribution of the shells generated through this mechanism
to be rather random, whereas the distribution of Magellanic Bridge
shells is confined to high column density regions.  The ram pressure
model can not be said to be a primary cause of Magellanic Bridge
expanding \hi\ shells.

\section{Summary}\label{sec:summary}
High resolution \hi\ maps of the inter-Magellanic Cloud region, the
Magellanic Bridge, reveal a complex and intricate structure of lumps,
filaments and holes across all observed scales, from \sm98\asec to
\sm7\deg. In general, much of the \hi\ of the Bridge appears to be
confined into two velocity components at 38\kms\ and 8\kms\ [VGSR]. This
bimodality converges to a single velocity of \sm23\kms\ [VGSR] at
\sm2\hr20\min, \sm3.6 kpc from the SMC towards the LMC. A large \hi\
loop, approximately 1 kpc in diameter, can be seen in the Bridge,
adjacent to the SMC.

A survey of \hi\ spherical expanding shells of the Magellanic Bridge,
has uncovered 163 examples of such structures.  Generally, shells
found within the Magellanic Bridge are less energetic, expand more
slowly and are smaller than those found within the SMC, although this
survey has shown that the mean kinematic age of shells in the
Magellanic Bridge is approximately equivalent to that of the SMC.
Although the \hi\ column density and OB distribution seem to spatially
correlate very well, as well as the distribution of \hi\ expanding
shell features and \hi\ column density, we have found very poor
correlation between the \hi\ shells and OB association distributions in
the Magellanic Bridge.  In addition, there appears to be a local {\em
  excess} of \hi\ immediately surrounding the positions of OB
associations.  These findings do not support the popular idea of
stellar wind being the driving engine of an \hi\ shell, although at
this time there are no alternative shell-generation mechanisms that
completely satisfactorily describe the energies and distribution of
the observed shell population. The distribution of incomplete shell-like
features of the Magellanic Bridge will be the focus of a future paper,
and will help to ascertain the plausability of the HVC model for shell
generation mechanisms.

A comparison with other \hi\ shell surveys of the Magellanic-type
galaxies, the SMC and \hoii, has shown that the survey appears to be
insensitive to shells with large radii.  An examination of other \hi\
surveys indicate that this survey used here was particularly rigorous
in the definition of an expanding shell. As a result of the strictness
of the selection criteria, we have found that a region of the Bridge,
in which an excess of older shells exists, corresponds with a region
of lower \hi\ velocity dispersion. On the basis of this, we have
suggested that shells are prone to fragmentation in a dynamic
environment, where the tendency to fragmentation is somehow dependent
on shell age and size. A future paper will focus on a census of
incomplete and fragmented shell-like structures.

The only known \ha\ shell in the Magellanic Bridge has been shown to be
defined also in \hi.  This is the only \hi\ shell that can be
unmistakably be attributed to a stellar driving engine, and apart from
it coincidence with the \ha\ region, does not appear to have any other
characteristics to distinguish it from the rest of the Magellanic
Bridge shell sample.

\section{Acknowledgements}
The Authors would like to extend their thanks to the Referee for
helpful advice and suggestions to improve this paper. Thanks also to
Dr. Grahame White, Dr.  Paul Jones, and also Dr. Raymond Haynes for
their assistance in the 1997 ATCA observations. Finally, the Authors
would like to acknowledge and thank the CSIRO and ATNF for time on the
ATCA and Parkes Telescopes. 

\section*{References}
\reference Ayal S., Piran T.,  2001, AJ, 555, 23
\reference Battinelli P., Demers S., 1998, AJ, 115, 1472 
\reference Battinelli P., Demers S., 1992, AJ, 104, 1458 
\reference Bica E.L.D., Schmitt H.R. 1995, ApJS, 101, 41 
\reference Blinnikov S.I., Novikov I.D., Perevodchikova T.V., Polnarev A.G,. 1984, SvA. 10, L177
\reference Brink, E., Bajaja E., 1986, A\&A 169, 14 
\reference Bureau M., Carignan C., 2002, AJ, 123, 1316
\reference Chevalier R.A., 1974, AJ, 188, 501 
\reference Demers S., Battinelli P., 1998, AJ 115, 154 
\reference Efremov Y.N., Elmegreen B.G., Hodge P.W., 1998, ApJ, 501, L163 
\reference Ehlerova S., Palous J., Theis Ch., Hensler G., 1997, A\&A, 328, 121
\reference Ehlerova S., Paulos J., 1996, A\&A, 313, 478 
\reference Gardiner L.T., Sawa T., Fujimoto F., 1994, MNRAS, 226, 567 
\reference Gardiner L.T., Noguchi M., 1996, MNRAS 278, 191
\reference Graham M.F., Smith R.J., Meaburn J., Bryce M., 2001, MNRAS 326, 539
\reference Grondin L., Demers S., 1993, in Sasselov D.D. ed, The
International Workshop on Luminous High-Latitude Stars,
Astron. Soc. Pac., San Francisco, p. 380 
\reference Grondin L., Demers S., Kunkel W.E., 1992, AJ, 103, 1234 
\reference Cox D.P., 1972, ApJ, 179, 159 
\reference Hindman J.V., McGee R.X., Carter A.W.L., Kerr F.J., 1961, AJ, 66, 45 
\reference Kim S., Dopita M.S., Staveley-Smith L., Bessell M.S., 1999, AJ, 118, 2797 
\reference Lozinskaya T.A., 1992, in Supernovae and Stellar Wind in the
Interstellar Medium, American Institute of Physics, New York, p.365
\reference Mathewson D.S., Cleary M.N., Murray J.D., 1974, APJ, 190, 291 
\reference McGee R.X., Newton L.M., 1986, Proc. ASA, 6/4, 471 
\reference Mathewson D.S., Ford V.L., 1984, in Van Den Bergh S., de Boer
K.S., eds, Proc. IAU Symp. 108, Structure and
evolution of the Magellanic Clouds, D. Reidel Pub. Co. Holland, p.125
\reference McClure-Griffiths N.M., Dickey J.M, Gaensler B.M., Green
A.J.,  Haynes R.F., Wieringa M.H.,2000, A.J, 119, 2828
\reference McCray R., Kafatos M., 1987, ApJ, 317, 190 
\reference McCray R., Snow T.P.Jr. 1979, ARA\&A, 17, 213 
\reference Meaburn,J. 1986, MNRAS, 223, 317 
\reference Oey M.S., Clarke C.J., 1997, MNRAS, 289, 570 
\reference Ott J.,  Walter F.,  Brinks E.,  Van Dyk S. D., Dirsch B.,
Klein U.,2001, AJ, 122 
\reference Paczy\'nski B., 1998, AJ, 494, L45 
\reference Parker Q. 1998, in Johnston H., Ricketts S., eds, AAO
Newsletter \#87, Anglo Australian Observatory, p.8
\reference Paturel G. et al., 1997, A\&AS, 124, 109
\reference Puche D., Westpfal D., Brinks E., Roy J-R, 1992, AJ, 103, 1841 
\reference Putman M.E., 2000, PASA, 17, 1 
\reference Putman M.E. et al., 1998, Nat., 394, 752
\reference Br\"{u}ns C., Kerp J., Staveley-Smith L. 2000, in
Kraan-Korteweg R.C., Henning P.A., Andernach H.,eds, IAU conf. proc. 218
Mapping the Hidden Universe: The Universe behind the Milky Way - The
Universe in HI.,Astron. Soc. Pac.,San Francisco, p.349
\reference Rhode, K.L., Salzer J.J., Wespfahl D.J., Radice L.A.,
1999, AJ, 118, 323
\reference Sawa T.; Fujimoto M.; Kumai Y. 1999, in Chu, Y-H., Suntzeff
N.B., Hesser J.E., Bohlender D.A., eds, IAU Symp. 190, New Views of
the Magellanic Clouds, Astron. Soc. Pac. San Francisco, p.499
\reference Shu F.H., 1992, in Osterbrock D.E., Miller, J.S.,eds, The
Physics of Astrophysics, University Science Books, Calif., p.264
\reference Shull J.M., Saken J.M., 1995, ApJ, 444, 663 
\reference Stanimirovi\'c S., Lazarian A., 2001, ApJ, 551, L53 
\reference Stanimirovi\'c S., 2000, AA Soc. 197, \#59.05 
\reference Stanimirovi\'c S., Staveley-Smith L., Dickey J.M., Sault
R.J., Snowden S.L., 1999, MNRAS, 301, 417 
\reference Staveley-Smith L., Kim S., Putman M., Stanimirovi\'c, S., 1998, in
Schielicke R.E.,ed, Rev. Mod. Astron. 11: Stars and Galaxies,
Astronomische Gesellschaft, Hamburg, p.117
\reference Staveley-Smith L., Sault R. J., Hatzidimitriou D., Kesteven
M.J., McConnell D., 1997, MNRAS, 289, 225 
\reference Tenorio-Tagle G., 1981, A\&A 94, 338 
\reference Tenorio-Tagle G. Franco J., Bodenheimer P., R\'o\.zyczka M. 1987,A\&A
179, 219 
\reference Tenorio-Tagle Bodenheimer P., R\'o\.zyczka M.,  Franco
J., 1986, A\&A, 170, 107 
\reference Vishniac E.T., 1983, AJ, 274, 152 
\reference Walter F., Brinks E., 2001, AJ, 121, 6 
\reference Weaver R., McCray R., Castor J., Shapiro P., Moore R.,
1977, ApJ, 218, 377 
\reference Wijers R.A.M.J., Bloom J.S., Bagla J.S., Natarajan P.,
1998, MNRAS, 294, L13 
reference Wilcots E.M., Miller A.J., 1998, AJ, 116, 2363 
\reference Zaritsky D., Harris J., Grebel E. K., Tompson I.B., 2000, APJ, 534, L53 

\bsp
\label{lastpage}
\end{document}